\documentclass[aps,pre,twocolumn,nobibnotes,superscriptaddress,longbibliography]{revtex4-2}

\usepackage{amssymb}
\usepackage{graphicx}
\usepackage{amsmath,amsfonts}
\usepackage{hyperref}
\usepackage[utf8] {inputenc}
\usepackage{xcolor}  
\usepackage[normalem]{ulem}
\usepackage[T1]{fontenc} 
\usepackage{multirow}
\usepackage{booktabs}
\usepackage{threeparttable}
\usepackage{tabularx}
\usepackage{enumitem}
\usepackage{float}
\hypersetup{
	colorlinks=true,
	allcolors=blue
}
\begin{document}
	\title{Robustness and size-dependence of circadian rhythms in multiscale suprachiasmatic-nucleus networks}

	\date{\today}
	\author{Youhao Zhuo}
	\affiliation{School of Physics and Electronic Engineering, Jiangsu University, Zhenjiang, Jiangsu, 212013, China}
	
	\author{Yingpeng Liu}
	\affiliation{School of Physics and Electronic Engineering, Jiangsu University, Zhenjiang, Jiangsu, 212013, China}
	
	\author{Jiao Wu}
	\affiliation{School of Mathematical Sciences, Jiangsu University, Zhenjiang, Jiangsu, 212013, China}
	
	\author{Kesheng Xu}
	\affiliation{School of Physics and Electronic Engineering, Jiangsu University, Zhenjiang, Jiangsu, 212013, China}
	
	\author{Muhua Zheng}
	\email[]{zhengmuhua163@gmail.com}
	\affiliation{School of Physics and Electronic Engineering, Jiangsu University, Zhenjiang, Jiangsu, 212013, China}
	
\begin{abstract}
Understanding how multi-scale network structure influences circadian rhythms in the suprachiasmatic nucleus (SCN) is essential for uncovering the principles of rhythmic robustness and synchronization. Previous studies using synthetic SCN networks suggested a size-dependent phenomenon, in which rhythmic activity initially strengthens with network size and then saturates, but it remains unclear whether this occurs in real SCN networks. Here, we apply geometric branch growth (GBG) and geometric renormalization (GR) to generate self-similar scaled-up and scaled-down replicas from a single-scale functional mouse SCN network. Unlike synthetic models, these SCN replicas do not exhibit size-dependent rhythms: average period, amplitude, and synchronization remain stable across scales. By increasing the average degree with network size, we reproduce size-dependent rhythms and show that they arise from network connectivity, whereas low-degree networks fragment and fail to sustain oscillations. Disrupting clustering self-similarity slightly reduces synchronization, but circadian rhythms remain robust, indicating that average degree, rather than clustering, is the dominant structural driver. These results highlight the resilience of SCN rhythms to network scaling and provide a framework for linking multi-scale network structure to biological timekeeping.
\end{abstract}
\maketitle

\section{Introduction}
From the opening of a flower at dawn to the sleep-wake patterns of humans, circadian rhythms shape life across plants, animals, and even microorganisms, synchronizing diverse behaviors and physiological processes with the natural light-dark cycle~\cite{welsh2010suprachiasmatic,aschoff1981survey,mendoza2025brain}.
Remarkably, animals that stay in continuous darkness still display free-running rhythms, and removal of the central circadian clock abolishes these rhythms, pointing to the existence of an endogenous biological timer~\cite{gu2019heterogeneity,zheng2022motif}. 
In mammals, the central circadian clock is located in the hypothalamic suprachiasmatic nucleus (SCN), which governs daily rhythms of behavior and physiology~\cite{pittendrigh1976,hastings2018,welsh2010}.
The SCN consists of roughly $20{,}000$ coupled neuronal oscillators, each modeled as a node in a complex network, with inter-neuronal couplings forming the links~\cite{welsh2010,welsh1995,gu2016,gu2021network,zhang2022hybrid,guo2022path,zheng2022few,feng2025model,yang2024synchronization,allard2024geometric,lin2024network}. 
Although its precise anatomical wiring remains unresolved, various network-based models—including all-to-all~\cite{gu2016}, nearest-neighbor~\cite{an2013neuropeptide,to2007molecular}, small-world~\cite{vasalou2009small}, random~\cite{gu2016}, and scale-free topologies~\cite{gu2016}—have been developed to explore how network structure maintains biological rhythms within the SCN.

Previous studies using mathematical models of synthetic SCN networks have revealed a size-dependent phenomenon in circadian rhythms: as the number of oscillators (i.e., network size) increases, synchronization initially strengthens and then saturates, while weakly rhythmic oscillators progressively regain their biological oscillations~\cite{BWnt}. Consistent with this, Vasalou et al.~\cite{vasalou2009small} also reported that increasing the network size contributes to improved biological rhythmicity. The number of oscillators is critical in maintaining SCN synchronization and rhythmicity.

However, whether size-dependent circadian rhythms occur in real SCN networks is still under debate. Most previous studies have examined either synthetic networks or empirical networks at a single scale, leaving the dynamics across multiple scales insufficiently characterized. Consequently, developing comprehensive multi-scale modeling frameworks is essential for elucidating the influence of network size on SCN synchronization and rhythmicity.
A key challenge is that, given only an empirical SCN network at a single scale, it remains unclear how to generate multi-scales networks while preserving the original structural features, thereby enabling the study of size-dependent biological rhythms.


In this work, we apply our previous geometric branch growth (GBG)~\cite{zheng2021scaling} and geometric renormalization (GR)~\cite{zheng2020geometric,Garcia2018} techniques to five functional SCN networks of mice~\cite{wu2023effects}, generating self-similar scaled-up and scaled-down replicas from a single-scale empirical network. GBG is a versatile method for producing scaled-up replicas, enabling the analysis of network behavior across different size scales. Together with the scaled-down replicas generated by GR~\cite{zheng2020geometric}, it enables a self-similar multiscale unfolding of complex networks across scales. Both GBG and GR build on the insight that real networks are organized according to an underlying hyperbolic geometry~\cite{Serrano2008,Krioukov2009}, where the probability of connections depends on node distances in this hidden space via a universal rule that applies across all scales. 
Our previous work has shown that SCN functional networks, as in many other domains, can be embedded in hidden hyperbolic spaces, where hyperbolic distance distinguishes short- and long-range connections~\cite{wu2023effects}. The inferred hyperbolic maps provide a biologically meaningful representation of network organization, revealing a spatial degree hierarchy with low-degree nodes in the shell and high-degree nodes in the SCN core~\cite{wu2023effects}. 
Moreover, the hyperbolic geometric framework captures key network features such as degree distributions and the small-world effect~\cite{Papadopoulos2012,Garcia2018aa,Zuev2015aa}. 
These findings support the idea that SCN networks follow the same underlying geometric principles observed in other complex systems and provide a foundation for applying the GBG and GR techniques.


With multi-scale replicas of the mouse SCN networks in hand, we investigate size-dependent circadian rhythms using the Becker--Weimann model, a molecular model of the SCN circadian system~\cite{BWnt}. Our results show that the period, amplitude, and synchronization ability of oscillations are preserved with increasing network size, remaining independent of network scale. 
However, increasing the average degree with the network  scale helps maintain rhythmicity and enhances synchronization between circadian oscillators, consistent with previous findings~\cite{BWnt,vasalou2009small}. Finally, we examine the impact of disrupting the self-similarity of SCN networks on synchronization.


The remainder of this paper proceeds as follows. Section~II describes the datasets for the functional SCN networks of mice, including the experimental sources and preprocessing methods. In Section~III, we introduce the GBG and GR methods together with the latent hyperbolic geometric models, detailing their theoretical foundation and implementation for generating multi-scale network replicas. Section~IV presents and analyzes the simulation results, with a focus on the emergence or absence of size-dependent circadian rhythms under different structural perturbations. Finally, Section~VI provides conclusions and discusses the broader implications of our findings for circadian rhythm regulation in complex networks.

\section{Datasets for Functional SCN Networks}
\begin{figure*}[t]
	\centering
	\includegraphics[width=1.0\linewidth]{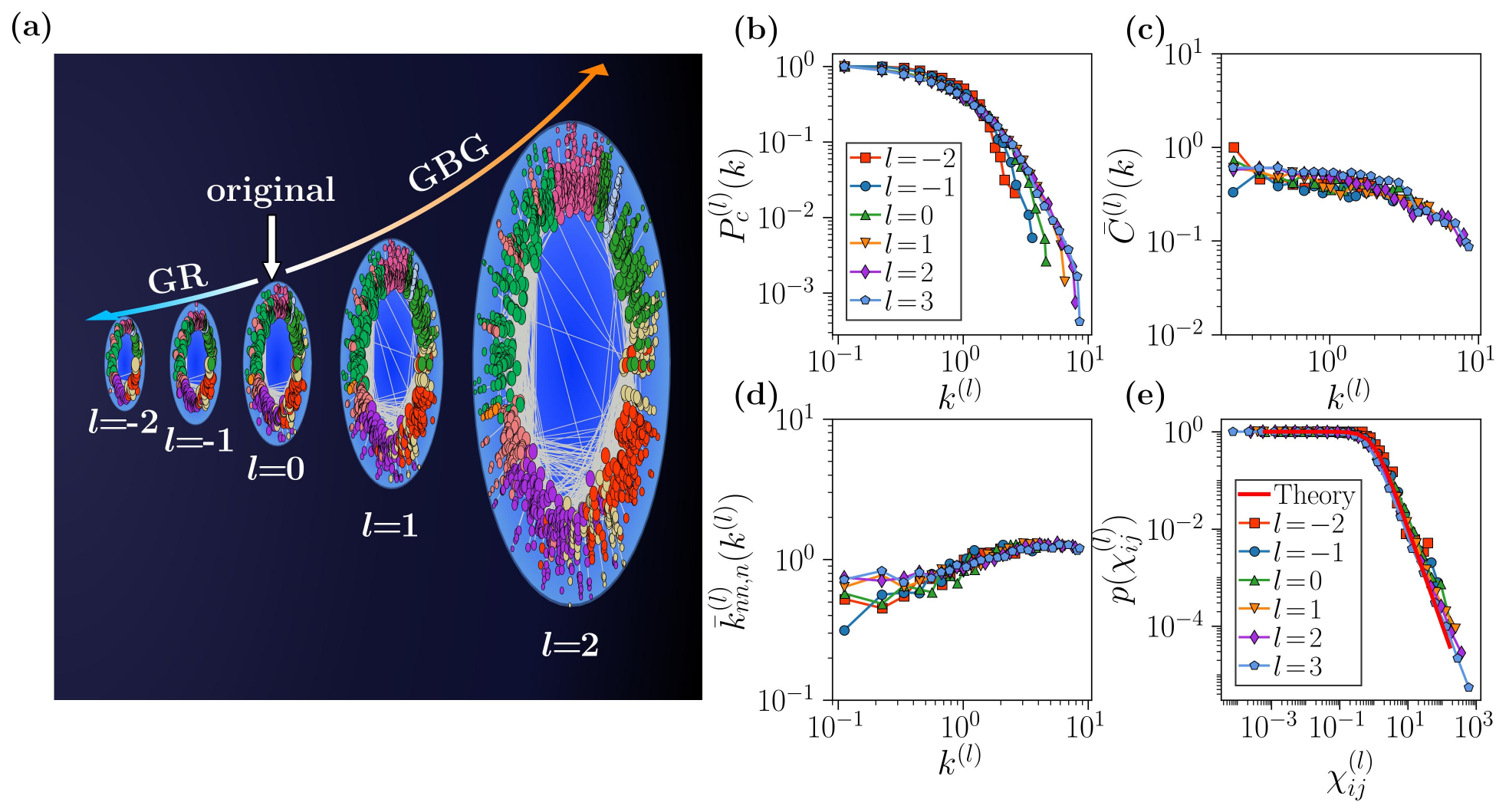}
\caption{
	{\bf Self-similarity of scaled-up and scaled-down SCN networks.} 
	(a) Schematic illustration of the GBG and GR models. Layer $l=0$ corresponds to the original SCN network. Different node colors indicate communities detected by the Louvain algorithm~\cite{Blondel2008}. Yellow and blue arrows represent the scaling-up and scaling-down processes of the multiscale unfolding under the GBG and GR models, respectively. For clarity, GR layers are relabeled as $l=-1,-2,\dots$ and GBG layers as $l=1,2,\dots$. In each layer, node size is proportional to $\log k$, where $k$ is the node degree, and node colors inherit those of their ancestors in the original network ($l=0$).  
	(b-d) Topological properties of scaled-up and scaled-down SCN networks.  
	(b) Complementary cumulative degree distribution. 
	(c) Degree-dependent clustering coefficient. 
	(d) Degree--degree correlations.  
	(e) Connection probability $p(\chi_{ij}^{(l)})$ as a function of the effective distance $\chi_{ij}^{(l)}$ in layer $l$. The red curve indicates the theoretical prediction from Eq.~\eqref{eq:con_pro_S1} for the original SCN network at $l=0$.
}
\label{fig:properties_non_isolated}
\end{figure*}

We analyze five functional networks of mouse SCN explants obtained from a TTX-mediated resynchronization experiment~\cite{abel2016functional} (all associated experimental data are available in Ref.~\cite{abel2016functional}). These networks were reconstructed from bioluminescence reporter data collected after TTX washout. During the recovery period, single-cell time-series data were analyzed using the maximal information coefficient (MIC) to infer functional connections.

Specifically, MIC values were computed for all pairs of raw bioluminescence traces from five independently cultured SCNs, without detrending or preprocessing. Because the SCNs were grown separately and do not interact, inferred links between different SCNs represent false positives, whereas links within the same SCN are potential true positives. Using this negative-control dataset, we constructed a receiver operating characteristic (ROC) curve to evaluate the ability of MIC to distinguish plausible biological connections from spurious ones. For each MIC threshold, the possible positive rate (PPR; within-SCN links) and false-positive rate (FPR; between-SCN links) were quantified.
This analysis identified a critical MIC threshold of $m_{\mathrm{crit}} = 0.935$, corresponding to an FPR of 0.0032 while retaining the strongest meaningful connections (PPR = 0.036). To account for small differences in synchronization across SCNs, the threshold for each SCN was slightly increased above $m_{\mathrm{crit}}$ to equalize the average node degree. The resulting thresholds are 0.949, 0.935, 0.990, 0.968, and 0.969 for SCNs~0-4, respectively. 

In each network, nodes represent individual cells, and edges are drawn between pairs of nodes whose MIC values exceed the corresponding threshold. For each SCN, only the largest connected component is retained, and the resulting networks are treated as undirected and unweighted. Table~\ref{tab:SCN_data} summarizes the main statistical properties of the resulting networks.

\begin{table}[!h] 
	\begin{ruledtabular}
	\centering		
	\caption{Statistical properties of the functional SCN networks. Columns list the network name, number of nodes ($N$), average degree ($\bar{k}$), average local clustering coefficient ($\bar{C}$), average path length ($\bar{l}$), the hyperbolic embedding parameters $\beta$ and $\mu$, and the stable distribution fitting parameters $[\alpha,\, \eta,\, c,\, d]$.}
	\label{tab:SCN_data}	
	\begin{tabular}{*{8}{l}}
		No.	&$N$ &$\bar{k}$  &$\bar{C}$ &$\bar{l}$ &$\beta$ &$\mu$ &$[\alpha,\, \eta,\, c,\, d]$\\
		\hline	
		SCN 0            &379    &8.87       &0.40       &4.95       &1.985      &0.036     &[1.1, 1.0, 53.2, 319.2] \\ 
		SCN 1            &228    &8.98       &0.37       &3.81       &1.789      &0.031      &[1.3, 1.0, 39.6, 108.5] \\ 
		SCN 2            &179    &7.88       &0.31       &4.10       &1.500      &0.026      &[1.7, 1.0, 26.3, 19.5] \\ 
		SCN 3            &155    &9.60       &0.32       &3.18       &1.411      &0.019      &[1.4, 1.0, 24.9, 47.1] \\ 
		SCN 4            &176    &9.12       &0.34       &3.46       &1.571      &0.025      &[1.3, 1.0, 30.2, 62.6] \\ 
	\end{tabular}
	\end{ruledtabular}
\end{table} 

\section{Method}

\subsection{Hyperbolic geometric models of complex networks}
\textbf{The $\mathbb{S}^1/\mathbb{H}^2$ model.}
The GBG and GR methods in this work are based on the $\mathbb{S}^1$ model~\cite{Serrano2008}, a hyperbolic geometric representation of complex networks. The $\mathbb{S}^1$ model can accurately reproduce many real-world networks' features within a one-dimensional circular space~\cite{Serrano2008}. In the model, each node $i$ is allocated two hidden variables $\kappa_i$ and $\theta_i$, representing its expected degree (popularity) and angular position (similarity), respectively. The $N$ nodes are placed uniformly on the circle, and the density is fixed to unity, so the radius is $R = N / 2\pi$.  
The connection probability $p_{ij}$ follows a gravity-law form: it increases with the product of popularities $\kappa_i \kappa_j$ and decreases with angular separation. Specifically,  
\begin{equation} \label{eq:con_pro_S1}
	p_{ij} = \frac{1}{1+\chi_{ij}^\beta},
\end{equation}
where $\chi_{ij} = \frac{R \Delta\theta_{ij}}{\mu \kappa_i \kappa_j}$ and $\Delta \theta_{ij} = \pi - |\pi - |\theta_i - \theta_j||$ is their angular distance. The model parameters are $\beta > 1$, controlling the average clustering $\bar{c}$, and $\mu$, tuning the average degree $\bar{k}$. 

The $\mathbb{S}^1$ model is equivalent to another geometric model, the $\mathbb{H}^2$ model~\cite{krioukov2009curvature}, where $\theta_i$ is preserved but $\kappa_i$ is mapped to a radial coordinate  
$
	r_i = R_{\mathbb{H}^2} - 2\ln\frac{\kappa_i}{\kappa_0},
$
with $R_{\mathbb{H}^2} = 2\ln\frac{N}{\pi\mu\kappa_0^2}$ the hyperbolic disk radius and $\kappa_0 = \min(\{\kappa_i\})$ the minimum expected degree. In this representation, Eq.~(\ref{eq:con_pro_S1}) becomes the Fermi-Dirac form  
\begin{equation} \label{eq:con_pro_H2}
	p_{ij} = \frac{1}{1+e^{\frac{\beta}{2}(H_{ij} - R_{\mathbb{H}^2})}},
\end{equation}
where the hyperbolic distance~\cite{krioukov2010hyperbolic}  
\begin{equation} \label{eq:Hij}
H_{ij}\!\!=\!\!\cosh^{-1}\!\!\left(\cosh r_i\cosh r_j\!-\!\sinh r_i\sinh r_j\cos\Delta\theta_{ij}\right)
\end{equation}
encodes both popularity and similarity.

\textbf{Inferring hyperbolic coordinates of SCN networks.}
Hyperbolic geometry has emerged as a powerful framework for describing the structure of complex networks.
Previous studies have shown that real network topology is better captured when Euclidean distances are supplemented with topological information such as node degree, clustering, and homophily~\cite{vertes2012simple,betzel2016generative}. In the hyperbolic framework, these features arise naturally: node degrees are encoded through Eq.~(\ref{eq:con_pro_S1}), while clustering and common-neighbor effects follow from the triangle inequality properties of hyperbolic space. It is therefore reasonable to expect that SCN networks, like many other real systems, follow similar hidden geometric principles.

Accordingly, each SCN network is embedded into hyperbolic space, where we infer the node coordinates along with the parameters $\beta$ and $\mu$ using Mercator~\cite{Garcia2019}. This tool takes the adjacency matrix 
($a_{ij} = a_{ji} = 1$ if nodes $i$ and $j$ are connected, and 0 otherwise) as input and infers the node coordinates and hyperbolic parameters by selecting the configuration that most likely generates the observed network under the $\mathbb{S}^1$ model.
Specifically, the hyperbolic coordinates are inferred by determining the sets $\{\kappa_i\}$ and $\{\theta_i\}$ that maximize  
\begin{equation}
	\mathcal{L} = \prod_{i<j} \big[p_{ij}\big]^{a_{ij}} \big[1 - p_{ij}\big]^{1 - a_{ij}},
\end{equation}
where $p_{ij}$ is the connection probability defined in Eq.~(\ref{eq:con_pro_S1}). 

Figure~\ref{fig:embedding} in the Appendix~\ref{SI-A} shows the topological validation of the embedding across all SCN networks, demonstrating that the $\mathbb{S}^1$ model accurately reproduces their topological features.
In addition, as shown in Fig.~\ref{fig:map} in the Appendix~\ref{SI-A}, the inferred hyperbolic structure reflects meaningful biological organization: the left and right SCN separate naturally in the hyperbolic map, and core and shell neurons are arranged from the center toward the periphery. Moreover, as reported in Ref.~\cite{wu2023effects}, hyperbolic distance naturally distinguishes short- and long-range connections. Together, these features suggest that hyperbolic embeddings may help identify core and shell neurons and characterize connectivity patterns in SCN networks.

\subsection{Upscaled SCN network replicas with GBG}
\label{GBG}
The GBG model, introduced in~\cite{zheng2021scaling}, provides a method for generating enlarged replicas of complex networks.
Scaled-up replicas are enlarged versions of real networks in which the number of nodes is increased while maintaining the topological features of the original network, particularly the average degree $\bar{ k}^{(0)}$ (see the schematic illustration of the GBG in Fig.~\ref{fig:properties_non_isolated}(a)). The GBG model enlarges networks by probabilistically splitting nodes~\cite{zheng2021scaling}. Each node generates $r$ descendants with probability $p$, yielding a scaled-up network of size $ N' = N\big(1 + p(r-1)\big) = bN,$
where $b=1+p$ denotes the branching rate. For simplicity, we consider the case $r=2$ and $p=1$, so that every ancestor node produces two descendants (i.e., $b=1+p=2$). 
To preserve constant node density, the radius of the similarity circle is adjusted as $R' = bR$. Once the hidden degrees $\kappa_i$, angular positions $\theta_i$ of all nodes, and the global parameters $\beta$ and $\mu$ are inferred in the original network, magnified replicas of the SCN network are generated as follows:

\textbf{(i) Assigning angular coordinates to descendants.} 
To preserve self-similar expansion, the ordering of nodes on the circle and their concentration within angular sectors (defining geometric communities~\cite{Garcia2018aa,Zuev2015aa}) must be maintained. 
For this purpose, descendants are placed at angular coordinates $\theta_i^{+}$ and $\theta_i^{-}$ to the left and right of their ancestor $i$, chosen uniformly within a small angular range $\Delta \theta^\pm$. 
We set $\Delta \theta^\pm = \min\{\tfrac{2\pi}{N'}, \tfrac{\Delta \theta_{ij}}{2}\}$, where $\Delta \theta_{ij} = \pi - |\pi - |\theta_i - \theta_j||$ is the angular distance between ancestor $i$ and its consecutive neighbor $j$ (on the left or right) in the parent layer.

\textbf{(ii) Assigning hidden degrees to descendants.}
To determine the hidden degrees $\kappa^{+}$ and $\kappa^{-}$ in the descendant layer, we require that the quantity $z = \kappa^\beta$ be conserved under GR, i.e., $z = z^{+} + z^{-}$. 
Moreover, the descendant hidden degrees are taken as independent and identically distributed random variables drawn from the same distribution as the ancestor layer, $\rho(\kappa)$ (or equivalently $\rho(z)$). Together, these constraints lead to
\begin{equation}\label{eq:global_constraint}
	\int \int \mathrm{d} z^{+}\mathrm{d} z^{-} \, \rho(z^{+}) \rho(z^{-}) 
	\delta \!\left( z - \left( z^{+} + z^{-} \right) \right) = \rho(z).
\end{equation}
Equation~\eqref{eq:global_constraint} requires $\rho(z)$ to be a stable distribution~\cite{levy1925,Borak2005,Nolan2018}, since the sum of two independent draws from $\rho(z)$ must have the same distribution up to scaling and location shifts. Stable laws are defined by four parameters $f(z;\alpha,\eta,c,d)$, with $\alpha \in (0,2]$ (tail exponent), $\eta \in [-1,1]$ (skewness), and $c,d$ controlling scale and location. 

Once the stable distribution parameters $(\alpha,\eta,c,d)$ of a given layer are known, the distribution for descendants follows directly from Eq.~(\ref{eq:global_constraint}). For $b=2$ (all nodes split), we obtain
\begin{equation}
	f(z^{\pm};\alpha^{\pm},\eta^{\pm},c^{\pm},d^{\pm}) = f(z^{\pm};\alpha,\eta,c/2^{1/\alpha},d/2),
\end{equation}
with invariant shape parameters $(\alpha,\eta)$ and rescaled $(c,d)$ ensuring that the ancestor distribution is recovered when $z^{+}$ and $z^{-}$ are summed. Using these functions and Bayes' rule, values of $z^{+}$ are sampled from the conditional probability $\rho(z^{+}|z)^{\mathrm{nor}}$, normalized to guarantee non-negativity. Specifically,
the hidden degree $z^+$ of a descendant is determined by the hidden degree $z$ of its ancestor. It is then computed numerically as
\begin{equation} \label{eq:generate_z_plus}
	\int_{z^{\pm}_{\mathrm{cut}}}^{z^+}\rho(z'_{+}|z)^{\mathrm{nor}} \textrm{d}z'_{+}=u, 
\end{equation}
where $u$ is a uniform random variable on $[0,1]$, and
\begin{equation} 
	\begin{aligned}
		\rho(z^{+}|z)^{\mathrm{nor}} &= C(z) \rho(z^{+}|z)\\
		&=  C(z) f(z^{+};\alpha^{\pm},\eta^{\pm}, c^{\pm}, d^{\pm})\\
		&\times f(z-z^{+};\alpha^{\pm},\eta^{\pm}, c^{\pm}, d^{\pm}).
	\end{aligned}
	\label{eq:z_condition}
\end{equation}
where $z^{\pm}_{\mathrm{cut}} = z_0 / 2^{1/\alpha}$, with $z_0 = \min(z)$ in the ancestor layer. $C(z)$ is the normalization constant in Eq.~\eqref{eq:z_condition}, given by
\begin{equation} \label{eq:Cz}
	\begin{aligned}
		C(z) = \Bigg[ &\int_{z^{\pm}_{\mathrm{cut}}}^{z - z^{\pm}_{\mathrm{cut}}} \mathrm{d}z^+ \, 
		f(z^{+};\alpha^{\pm},\eta^{\pm}, c^{\pm}, d^{\pm}) \\
		&\times f(z - z^{+};\alpha^{\pm},\eta^{\pm}, c^{\pm}, d^{\pm}) \Bigg]^{-1}.
	\end{aligned}
\end{equation}

The complementary value $z^{-}$ is then computed as $z^{-} = z - z^{+}$, and both quantities are converted to hidden degrees using $\kappa = z^{1/\beta}$.

\textbf{(iii) Connections in the descendant layer.}
After the coordinates are determined for the descendant nodes, links are established to ensure that the resulting network belongs to the $\mathbb{S}^1$ ensemble. In the {\em non-inflationary} limit, GR consistency requires that only descendants of the same ancestor or of connected ancestors may connect. 
Using the $\mathbb{S}^1$ probability $p_{ij}$~\eqref{eq:con_pro_S1} with rescaled $\mu' = b\mu$ (keeping $\beta$ invariant), we connect descendants of the same ancestor and, for connected ancestors, allow descendant-descendant links with probability $p_{ij}(\mu')$, ensuring at least one such link.

In the {\em inflationary} limit, we rescale $\mu' \to a\mu'$ with $a>1$ and establish unbiased extra links to reach the target average degree. 
Equivalently, starting from the non-inflationary GBG map, we set $\mu'_a = ab\mu$ ($a \geq 1$) and connect previously unlinked pairs with probability
\begin{equation}\label{eq:conprob_extra}
	\pi_{ij} = \frac{p_{ij}(\mu'_a)-p_{ij}(\mu')}{1-p_{ij}(\mu')}. 
\end{equation}
Given the target size, $b$ and $l$ are fixed. To tune the average degree, we set 
$a=\varepsilon \bar{k}^{(0)} / \bar{k}^{(l)}$, 
where $\bar{k}^{(0)}$ is the target and $\bar{k}^{(l)}$ the non-inflationary value at layer $l$. Starting with $\varepsilon=1$, we iteratively add links using Eq.~\eqref{eq:conprob_extra} and compute $\bar{k}_{\mathrm{new}}^{(l)}$. If $\bar{k}_{\mathrm{new}}^{(l)} > \bar{k}^{(0)}$, the realization is discarded and $\varepsilon \to \varepsilon - 0.1u$; if $\bar{k}_{\mathrm{new}}^{(l)} < \bar{k}^{(0)}$, then $\varepsilon \to \varepsilon + 0.1u$, with $u \sim U(0,1)$. The process stops once $|\bar{k}_{\mathrm{new}}^{(l)} - \bar{k}^{(0)}| < 0.05$.

\subsection{Scaled-down SCN network replicas with GR}
\label{GR}
The GR transformation can generate a series of scaled-down replicas from a single scale network (see the schematic illustration of the GR in Fig.~\ref{fig:properties_non_isolated}(a)). This transformation coarsens a network by grouping adjacent nodes on the similarity circle into non-overlapping blocks of size $r$~\cite{zheng2020geometric}. Each block is merged into a supernode: two supernodes are connected if any nodes within their blocks are connected. Repeating this generates a hierarchy of self-similar, multi-scale networks.

Starting from layer $l=0$ with $r=2$, each supernode inherits an angular coordinate preserving the local order. Hidden degrees and positions are updated via
\begin{equation}
	\kappa' = \Big[ \sum_{j=1}^{r} (\kappa_j)^\beta \Big]^{1/\beta}, \quad
	\theta' = \Big[ \frac{\sum_{j=1}^{r} (\theta_j \kappa_j)^\beta}{\sum_{j=1}^{r} (\kappa_j)^\beta} \Big]^{1/\beta}.
\end{equation}
Global parameters are rescaled as $\mu' = \mu / r$, $\beta' = \beta$, $R' = R / r$. 

After obtaining a renormalized network via GR, the average degree must be reduced to match the original network. Following~\cite{Garcia2018}, we adjust $\mu_{\mathrm{new}}^{(l)}$ so that $\bar{k}_{\mathrm{new}}^{(l)} = \bar{k}^{(0)}$ using 
$\mu_{\mathrm{new}}^{(l)} = \varepsilon \, \bar{k}^{(0)} / \bar{k}^{(l)} \, \mu^{(l)}$, with $\varepsilon = 1$ initially.  
For each $\varepsilon$, the network is pruned: if $\bar{k}_{\mathrm{new}}^{(l)} > \bar{k}^{(0)}$, we update $\varepsilon \to \varepsilon - 0.1 u$; if $\bar{k}_{\mathrm{new}}^{(l)} < \bar{k}^{(0)}$, then $\varepsilon \to \varepsilon + 0.1 u$, where $u \sim U(0,1)$. 
The process stops once $|\bar{k}_{\mathrm{new}}^{(l)} - \bar{k}^{(0)}| < 0.05$.

\subsection{The Becker-Weimann Model}

We adopt the Becker--Weimann model to study size-dependent circadian rhythms~\cite{BWnt}. The intracellular oscillator comprises seven variables representing clock gene mRNAs and proteins, forming interlocked transcriptional/translational feedback loops that capture essential mammalian circadian dynamics. Intercellular coupling among SCN cells is mediated by circadian release of a neuropeptide~\cite{liu2000gaba, reed2001vasoactive, maywood2006synchronization}, assumed to be triggered by the cytosolic PER/CRY complex~\cite{BWnt}. This release activates a two-step cascade in connected cells via PKA and CREB, driving Per/Cry transcription. Including the neurotransmitter and cascade, the full system has ten variables: $Y_1$ ($Per/Cry$ mRNA), $Y_2$ (PER/CRY cytosolic complex), $Y_3$ (nuclear PER/CRY complex), $Y_4$ ($Bmal1$ mRNA), $Y_5$ (cytoplasmic BMAL1), $Y_6$ (nuclear BMAL1), $Y_7$ (active BMAL1*), $V$ (neurotransmitter), $X_1$ (PKA), and $X_2$ (CREB). For each node $i=1,\ldots,N$, the dynamics of these ten variables under constant darkness are governed by:

\begin{subequations}\label{eq:system}
	\renewcommand{\theequation}{\theparentequation-\alph{equation}}
	\begin{align}
		\frac{dY_{1,i}}{dt} &= f_{\text{Per/Cry},i} - k_{1d}Y_{1,i}+L , \label{eq:Y_1} \\
		\frac{dY_{2,i}}{dt} &= k_{2b}Y_{1,i}^q - (k_{2d}\!+\!k_{2t})Y_{2,i} + k_{3t}Y_{3,i}, \label{eq:Y_2} \\
		\frac{dY_{3,i}}{dt} &= k_{2t}Y_{2,i} - (k_{3t}\!+\!k_{3d})Y_{3,i}, \label{eq:Y_3} \\
		\frac{dY_{4,i}}{dt} &= f_{\text{Bmal},i} - k_{4d}Y_{4,i}, \label{eq:Y_4} \\
		\frac{dY_{5,i}}{dt} &= k_{5b}Y_{4,i} - (k_{5d}\!+\!k_{5t})Y_{5,i} + k_{6t}Y_{6,i}, \label{eq:Y_5} \\
		\frac{dY_{6,i}}{dt} &= k_{5t}Y_{5,i} - (k_{6d}\!+\!k_{6t}\!+\!k_{6a})Y_{6,i} + k_{7a}Y_{7,i}, \label{eq:Y_6} \\
		\frac{dY_{7,i}}{dt} &= k_{6a}Y_{6,i} - (k_{7d}\!+\!k_{7a})Y_{7,i}, \label{eq:Y_7} \\
		\frac{dV_i}{dt} &= k_8Y_{2,i} - k_{8d}V_i, \label{eq:V} \\		
		\frac{dX_{1,i}}{dt} &= k_{x1}Q_i(X_{1T}\!-\!X_{1,i}) - k_{dx1}X_{1,i}, \label{eq:X_1} \\
		\frac{dX_{2,i}}{dt} &= k_{x2}X_{1,i}(X_{2T}\!-\!X_{2,i}) - k_{dx2}X_{2,i}. \label{eq:X_2}
	\end{align}
\end{subequations}

The nonlinear transcription functions in Eq.~\eqref{eq:Y_1} and~\eqref{eq:Y_4} are 
\begin{align}
	f_{\text{Per/Cry},i} &= v_{1b} \frac{Y_{7,i} + X_{2,i}^h}{k_{1b} \left( 1 + \left( \frac{Y_{3,i}}{k_{1i}} \right)^p \right) + (Y_{7,i} + X_{2,i}^h)}, \nonumber \\
	f_{\text{Bmall},i} &= v_{4b} \frac{Y_{3,i}^r}{k_{4b}^r + Y_{3,i}^r}. \nonumber
\end{align}

The coupling term $Q_i$ in Eq.~\eqref{eq:X_1} represents a signaling cascade that activates the \textit{Per/Cry} promoter and is proportional to the local mean field $F_i$, given by  
\begin{equation}
	\label{eq:Q}
	Q_i = K F_i = K \sum_{j=1}^{N} \frac{a_{ij} V_j}{k_i},
\end{equation}
where $K$ denotes the coupling strength, $a_{ij}$ the entries of the adjacency matrix, and $k_i=\sum_{j=1}^{N} a_{ij}$ the degree of node $i$. 

We simulate light entrainment by a clipped sine wave,
\begin{equation}
	L(t)\!=\!
	\begin{cases}
		L_0 \sin\!\left(\dfrac{\pi t}{t_{\text{light}}}\right),\!\!
		& \! \text{if } t \bmod (t_{\text{light}} \!+\! t_{\text{dark}}) \! \le \! t_{\text{light}},\\ 
		0, & \! \text{otherwise}.
	\end{cases}
\end{equation}
where $L_0$ denotes the light intensity, i.e., the maximum amplitude of the light input. This formulation produces a periodic light--dark signal in which $0 < L(t) \le L_0$ during the light phase and $L(t)=0$ during the dark phase, alternating every $t_{\text{light}}=12$ and $t_{\text{dark}}=12$ hours.
Unless otherwise specified, we consider only the constant-darkness condition, i.e., $L(t)=0$.

To introduce heterogeneous periods, each system variable is scaled by a factor $e_i = 1/u_i$ with $u_i$ sampled from a Gaussian distribution (mean 1, standard deviation 0.05). This yields a period distribution between $20$ and $28$ hours.

In this paper, the Becker--Weimann model is simulated with a fourth-order Runge--Kutta method (time step 0.01 h), initializing variables from a normal distribution (mean 1, standard deviation 0.05). The first $100$ h are discarded as transient dynamics, and statistical analysis is performed on the following $100$ h of the time series. The parameters are set as in Ref.~\cite{BWnt}: 
$v_{1b} = 9.0 \, \text{nMh}^{-1}$, $ k_{1b} = 1.0 \, \text{nM}$, $ k_{1i} = 0.56 \, \text{nM}$, $ p = 3$, $ h = 2$, $ k_{1d} = 0.18 \, \text{h}^{-1},k_{2b} = 0.3 \, \text{h}^{-1}\, \text{nM}^{-(q-1)}$, $ q = 2$, $ k_{2d} = 0.1 \, \text{h}^{-1}$, $ k_{2t} = 0.36 \, \text{h}^{-1}$, $ k_{3t} = 0.02 \, \text{h}^{-1}$, $ k_{3d} = 0.18 \, \text{h}^{-1},v_{4b} = 1.0 \, \text{nM}$, $ k_{4b} = 2.16 \, \text{h}^{-1}$, $ r = 3$, $ k_{4d} = 1.1 \, \text{h}^{-1}$, $ k_{5b} = 0.24 \, \text{h}^{-1},k_{5d} = 0.09 \, \text{h}^{-1}$, $ k_{5t} = 0.45 \, \text{h}^{-1}$, $ k_{6t} = 0.06 \, \text{h}^{-1}$, $ k_{6d} = 0.18 \, \text{h}^{-1}$, $ k_{6a} = 0.09 \, \text{h}^{-1},k_{7a} = 0.003 \, \text{h}^{-1}$, $ k_{7d} = 0.13 \, \text{h}^{-1}$, $ k_{8} = 1.0 \, \text{h}^{-1}$, $ k_{8d} = 4.0 \, \text{h}^{-1},k_{x1} = 3.0 \, \text{h}^{-1} \, \text{nM}^{-1}$, $ X_{1T} = 15.0 \, \text{nM}$, $ k_{dx1} = 4.0 \, \text{h}^{-1},k_{x2} = 0.25 \, \text{h}^{-1} \, \text{nM}^{-1}$, $ X_{2T} = 15.0 \, \text{nM}$, $ k_{dx2} = 10.0 \, \text{h}^{-1}$. 
 
The dynamics of individual neuronal oscillators are represented by the Per/Cry mRNA concentration $Y_{1,i}$, and the SCN network by its average output
\begin{equation}
	\overline{Y_1} = \frac{1}{N} \sum_{i=1}^{N} Y_{1,i}.
\end{equation}
In our work, $Y_{1,i}$ is used as a marker of the dynamical evolution of individual neuronal oscillators. Accordingly, circadian rhythm strength is quantified based on $Y_{1,i}$ using three measures: the average period $T$, the SCN amplitude $A$, and the synchronization degree $R$. The average period is
\begin{equation}
	T = \frac{1}{N} \sum_{i=1}^{N} T_i,
\end{equation}
where $T_i$ is the period of $Y_{1,i}$. Note that oscillators with zero amplitude correspond to a steady (non-oscillatory) state, commonly referred to as oscillation death. In this case, oscillator $i$ does not exhibit sustained rhythmic behavior, and the period $T_i$ is not well defined; for convenience, we set $T_i = 0$. The average amplitude is defined as
\begin{equation}
	A = \frac{1}{N} \sum_{i=1}^{N} A_i,
\end{equation}
where $A_i$ denotes the amplitude of $Y_{1,i}$ for the $i$-th neuronal oscillator. Synchronization among oscillators is quantified by the order parameter $R$, defined as
\begin{equation}
	\label{eq:R}
	R = \frac{\langle \overline{Y_1}^2 \rangle - \langle \overline{Y_1} \rangle^2}{\frac{1}{N} \sum_{i=1}^{N} \left(\langle Y_{1,i}^2 \rangle - \langle Y_{1,i} \rangle^2 \right)}, 
\end{equation}
where $\langle \cdot \rangle$ denotes a time average. 

Note that the variability in the period $T$, amplitude $A$, and synchronization degree $R$ arises from both heterogeneity in $\mu_i$ and differences in the initial conditions, with heterogeneity in $\mu_i$ being the dominant source. Accordingly, all reported results are averaged over multiple realizations to account for residual variability.

\section{Results}

\subsection{Robustness of circadian rhythms in self-similar scaled-up and scaled-down SCN networks}

We apply the GBG and GR transformations to the five SCN networks (see the schematic illustration of the GBG and GR models in Fig.~\ref{fig:properties_non_isolated}(a)) and observe clear statistical self-similarity in Figs.~\ref{fig:properties_non_isolated}(b-e), where the curves from networks of different sizes naturally overlap in the representative SCN~0 network (see additional examples in Fig.~\ref{s2} in Appendix~\ref{SI-D}). Although no universal metric exists for quantifying self-similarity, it can be qualitatively assessed by the collapse of network-statistic curves across layers, with such overlap indicating scale-invariant self-similar behavior.

For each layer, we analyze the following structural properties: the complementary cumulative degree distribution, the degree-dependent clustering coefficient, and the degree--degree correlations characterized by the normalized nearest-neighbor degree,  
$
	\bar{k}_{nn,n}(k) = \bar{k}_{nn}(k)\,\frac{\overline{k}}{\overline{k^2}}.
$
Since the scaled-up and scaled-down SCN replicas preserve the average degree of the original network, these metrics are not rescaled by $\bar{k}$. In addition, we examine the connection probability $p(\chi_{ij}^{(l)})$ as a function of the effective distance $\chi_{ij}^{(l)}$ in each layer $l$. The red curve in Fig.~\ref{fig:properties_non_isolated}(e) shows the theoretical prediction from Eq.~\eqref{eq:con_pro_S1} for the original SCN network ($l=0$), indicating that node interactions are governed by distances in the latent space via a universal connectivity law that captures both short- and long-range links across scales. Across all panels in Figs.~\ref{fig:properties_non_isolated}(b-e), the results exhibit pronounced self-similar features, with curves from distinct networks collapsing onto a single profile. 
These observations demonstrate that SCN networks are self-similar and confirm that the GBG and GR transformations provide a general framework for multi-scale analysis of complex networks.

Moreover, we use the Kolmogorov--Smirnov (KS) distance to quantify the self-similarity of the degree distributions~\cite{simard2011computing}. A small KS distance together with a large $p$-value indicates a higher degree of self-similarity relative to the original network. As shown in Table~\ref{tab:KS_test} in Appendix~\ref{KStest}, aside from finite-size effects in the smallest network, the degree distributions remain statistically consistent with those of the original network, further supporting robust self-similarity and structural preservation across scales.

\begin{figure}[t]
	\centering
\includegraphics[width=1.0\linewidth]{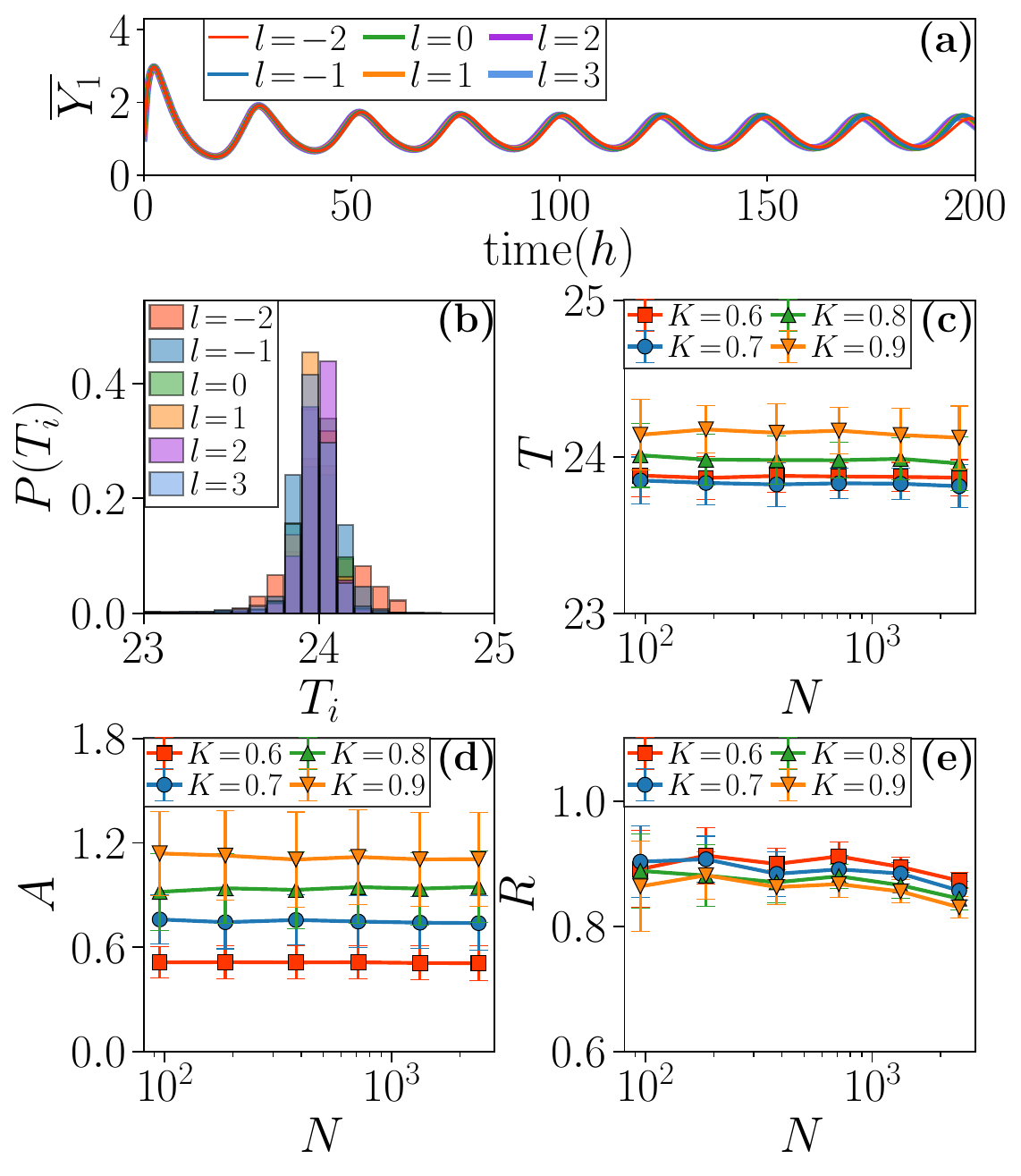}
\caption{
	\textbf{Circadian rhythms in scaled-up and scaled-down SCN replicas.}
	(a) Time series of the average output $\overline{Y_1}$ across different layers $l$ for coupling strength $K = 0.8$.
	(b) Corresponding period distribution $P(T_i)$ of individual oscillators for $K = 0.8$.
	(c) Average period $T$, (d) SCN amplitude $A$, and (e) synchronization degree $R$ as functions of network size $N$ for different coupling strengths.
	Error bars represent one standard deviation from the mean. Results are obtained using the Becker--Weimann model under constant-darkness conditions and averaged over $50$ realizations with different initial conditions and independently sampled $\mu_i$.
}
\label{fig:Fig2}
\end{figure}

Next, we apply the Becker--Weimann model under constant-darkness condition to upscaled and downscaled SCN replicas to examine circadian rhythms across networks of different sizes. Because weak coupling leads to oscillation death (see Fig.~\ref{fig:K_check} of Appendix~\ref{SI-D}), thereby suppressing coherent circadian oscillations and disrupting biologically meaningful rhythmic activity, we focus on the strong-coupling regime across scales. As shown in Fig.~\ref{fig:Fig2}, the rhythms are not affected by network size. Specifically, Fig.~\ref{fig:Fig2}(a) presents the time series of the average output $\overline{Y_1}$ in representative SCN~0 replicas for different layers $l$ with coupling strength $K=0.8$. The curves nearly coincide, visually confirming that the average circadian rhythm is robust to scaling. Both the oscillation period and amplitude remain unchanged regardless of whether the network is expanded or reduced. Furthermore, the narrow and overlapping distributions of individual oscillator periods $P(T_i)$ across layers demonstrate that the circadian period is highly conserved, reinforcing the scale-invariant behavior of SCN networks (see Fig.~\ref{fig:Fig2}(b)).

To further clarify the above results, as shown in Figs.~\ref{fig:Fig2}(c-e), we examine the average period $T$, the SCN amplitude $A$, and the synchronization order parameter $R$ as functions of network size $N$ under different coupling strengths. 
Results for the remaining SCN networks are presented in Fig.~\ref{s3} of Appendix~\ref{SI-D}. 
For a fixed coupling strength, the average period, amplitude, and synchronization degree remain nearly unchanged as the network scale increases. Strong synchronization is consistently observed with $R>0.80$ across all scales in Fig.~\ref{fig:Fig2}(e). Strikingly, these scale-invariant SCN rhythms are robust both under light--dark cycles and within the Kuramoto oscillator model (see Figs.~\ref{fig:light_dark} and~\ref{fig:Kuramoto} in Appendix~\ref{light-dark}). 
Taken together, these findings demonstrate that circadian rhythms in self-similar SCN replicas are robust and largely independent of network size.

\subsection{Emergence of size-dependent circadian rhythms driven by average degree} 
\begin{figure*}[ht]
	\centering
	\includegraphics[width=1.0\linewidth]{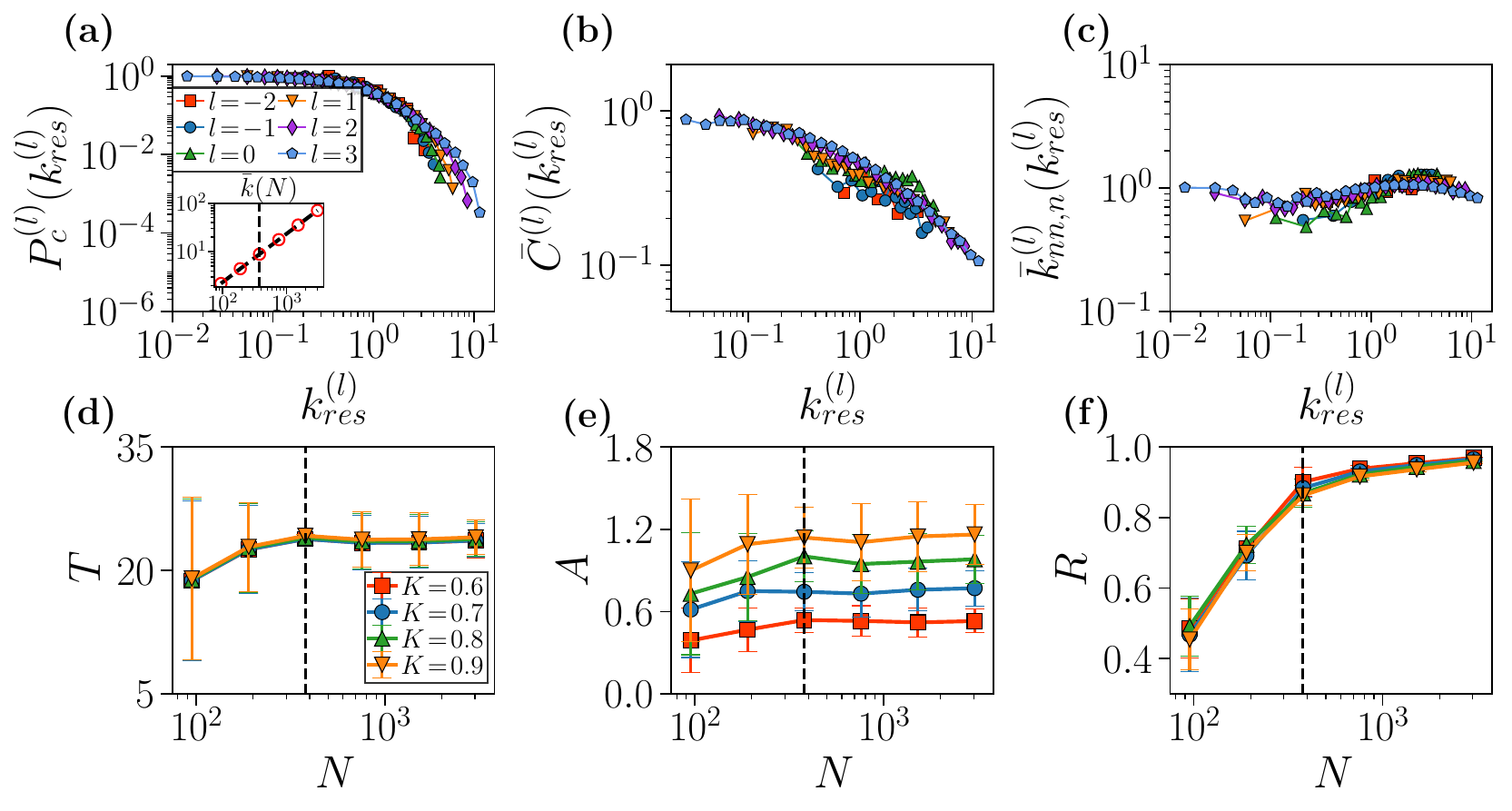}
	\caption{
		\textbf{Effects of average degree on circadian rhythms.}  
		(a) Complementary cumulative degree distribution; inset: average degree as a function of network size $N$, showing the expected increase with $N$.    
		(b) Degree-dependent clustering coefficient, 
		(c) Degree--degree correlations.  
		In (a-c), degrees are rescaled as $k_{\mathrm{res}} = k / \bar k$.  
		(d) Average period $T$, (e) SCN amplitude $A$, and (f) synchronization degree $R$ as functions of $N$ in the perturbed networks from Null-$k$ model under varying coupling strengths. Error bars represent one standard deviation from the mean. The black dashed vertical lines indicate the original networks.  
		Results are averaged over $50$ realizations with different initial conditions and independently sampled $\mu_i$.
	}
	\label{fig:3}
\end{figure*}

\begin{figure}[!ht]
	\centering
	\includegraphics[width=1.0\linewidth]{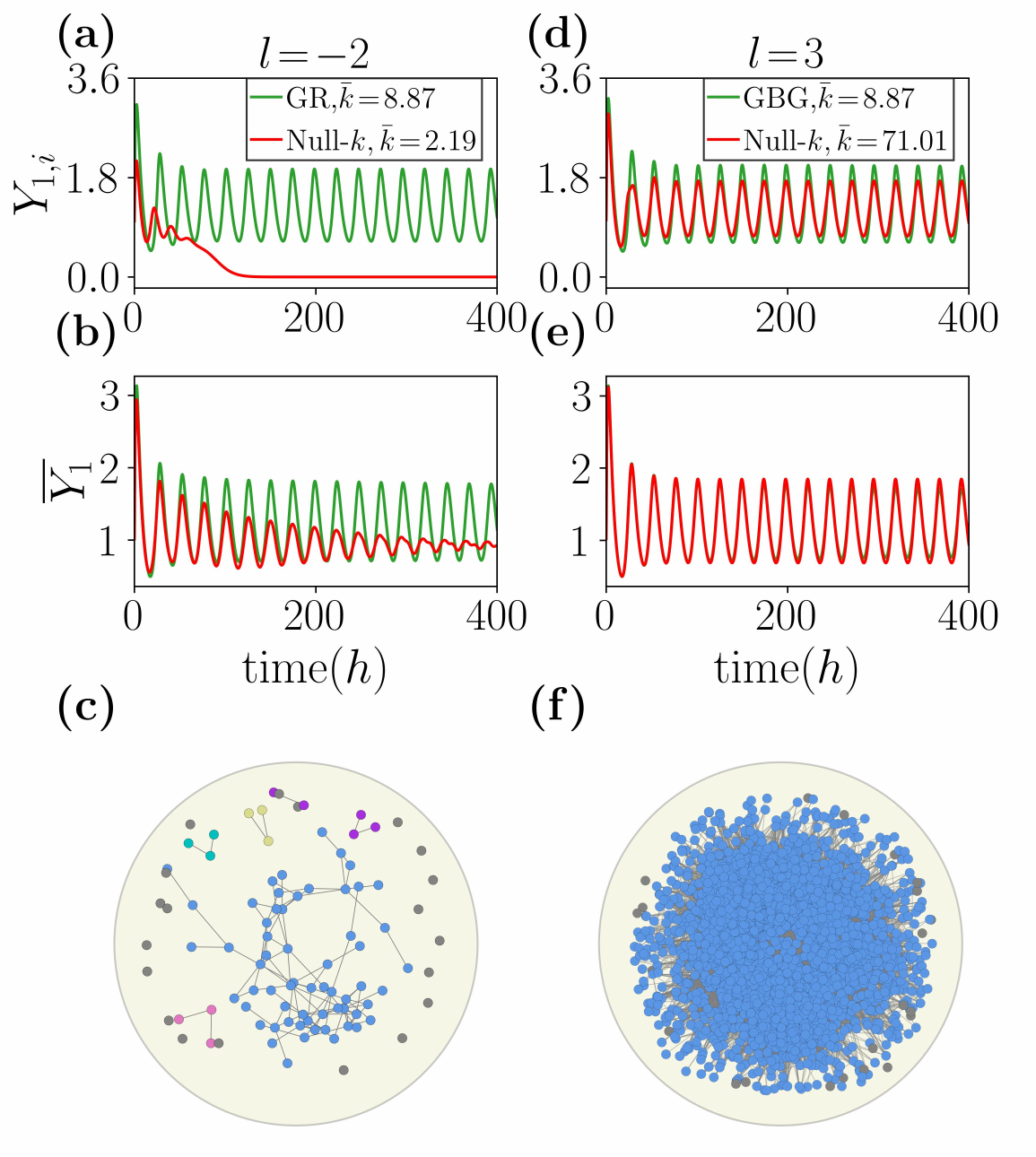}
\caption{
	\textbf{Evolution of neuronal oscillators in networks with different average degrees.}  
	(a, b) Time series of a representative neuronal oscillator $Y_{1,i}$ and the population-averaged output $\overline{Y_1}$ in layer $l\!=\!-2$ from the Null-$k$ model and the GR replica, respectively.  
	(c) Corresponding network snapshot from the Null-$k$ model.  
	(d, e) Time series of a representative neuronal oscillator $Y_{1,i}$ and the population-averaged output $\overline{Y_1}$ in layer $l\!=\!3$ from the Null-$k$ model and the GBG replica, respectively.  
	(f) Corresponding network snapshot from the Null-$k$ model.  
	The GR replica at layer $l=-2$ and the GBG replica at layer $l=3$ are used as references for comparison.
}
\label{fig:4}
\end{figure}

The discrepancies between our findings and previous studies on size-dependent circadian rhythms~\cite{BWnt} motivated closer scrutiny. In Ref.~\cite{BWnt}, synchronization is examined in networks of varying sizes while keeping the connectivity $c_0$ fixed, where $c_0$ is defined as the ratio of actual to maximal possible connections. Under this assumption, the average in-degree $\overline{k}_{\mathrm{in}} = c_0 (N-1)$ increases with network size $N$, resulting in progressively stronger coupling and enhanced synchronization. In contrast, we preserve the average degree as a biologically relevant parameter reflecting network density and effective coupling, thereby providing a meaningful baseline for interaction strength and clarifying how increasing degree with system size can enhance synchronization.

To examine the effects of average degree on circadian rhythms, we propose a \emph{Null-$k$} model that increases the average degree with network size by modifying the GR and GBG methods (see Appendix~\ref{SI-B} for more details). Specifically, the original network at $l=0$ remains unchanged. In the GR layers ($l=-1,-2,\dots$), we reduce the targeted average degree by pruning edges, while in the GBG layers ($l=-1,-2,\dots$), we increase the targeted average degree by adding additional edges. 
As an illustrative example, Figs.~\ref{fig:3}(a-c) summarize the structural properties of the resulting networks under a simple artificial scaling rule, $\bar{k}^{(l)} = 2 \, \bar{k}^{(l-1)},$
where the mean degree doubles at each layer.
As expected, the Null-$k$ model increases the average degree with network size $N$ (see Fig.~\ref{fig:3}(a), inset) while preserving key structural properties, including the complementary cumulative degree distribution, degree-dependent clustering coefficient, and degree--degree correlations (see Figs.~\ref{fig:3}(a-c)). These metrics are evaluated with respect to the rescaled degree $k_{\rm res} = k/\bar{k}$, normalizing for differences in average degree across networks. In Figs.~\ref{fig:3}(a-c), pronounced self-similar features are evident, as curves from networks of different sizes collapse onto a single profile when degrees are rescaled by the corresponding average degree.

Interestingly, in networks generated by the Null-$k$ model, we observe size-dependent circadian rhythms: the average period, amplitude, and synchronization degree increase with network size initially and then saturate (see Figs.~\ref{fig:3}(d-f) for a representative SCN~0, and the remaining networks in Fig.~\ref{s4} of Appendix~\ref{SI-B}).
Compared with the original network at $l=0$ (black dashed vertical lines in Figs.~\ref{fig:3}(d-f)), circadian rhythms in smaller networks with lower average degree are more sensitive and exhibit higher variance. The large variability of $T$, $A$, and $R$ indicates poor, non-robust, network-dependent rhythms, implying that the observed size dependence is driven by average degree. Note that synchronization and amplitude exhibit different sensitivities to the coupling strength in Fig.~\ref{fig:3}(d-f) and Fig.~\ref{fig:Fig2}(c-e). Once $K$ exceeds the phase-locking threshold, synchronization saturates because global phase coherence is already established, whereas the oscillation amplitude $A$ continues to increase since stronger coupling further suppresses phase dispersion and destructive interference (see Fig.~\ref{fig:K_check} in Appendix~\ref{SI-D} for a detailed analysis).

%


To further investigate the origin of size dependence, we examine the time evolution of neuronal oscillators in the smallest ($l=-2$) and largest ($l=3$) networks. Figs.~\ref{fig:4}(a,b) show the time series of a representative oscillator $Y_{1,i}$ and the average output $\overline{Y_1}$ in layer $l=-2$ for the Null-$k$ model and GR replica, respectively. In Fig.~\ref{fig:4}(a), an isolated oscillator fails to sustain oscillations due to the loss of interactions with other nodes, whereas in the GR replica the same node maintains oscillations through coupling with its neighbors. Representative average outputs $\overline{Y_1}$ in Fig.~\ref{fig:4}(b) for the Null-$k$ network are damped or irregular compared with those in the GR replica. We also find that the Null-$k$ network is fragmented into many components and isolated nodes (see snapshot in Fig.~\ref{fig:4}(c)), preventing oscillators from sustaining oscillations and resulting in poor circadian rhythms. 
However, in the largest network ($l=3$) with very large average degree, oscillations are sustained as in the GR replica, both for the representative oscillator $Y_{1,i}$ (Fig.~\ref{fig:4}(d)) and the average output $\overline{Y_1}$ (Fig.~\ref{fig:4}(e)). As shown in Fig.~\ref{fig:4}(f), this network is well connected, with a giant component comprising $99.3\%$ of the nodes.
These results confirm that size-dependent circadian rhythms arise from the increase in average degree with network size.

\begin{figure*}[ht]
	\centering
	\includegraphics[width=1.0\linewidth]{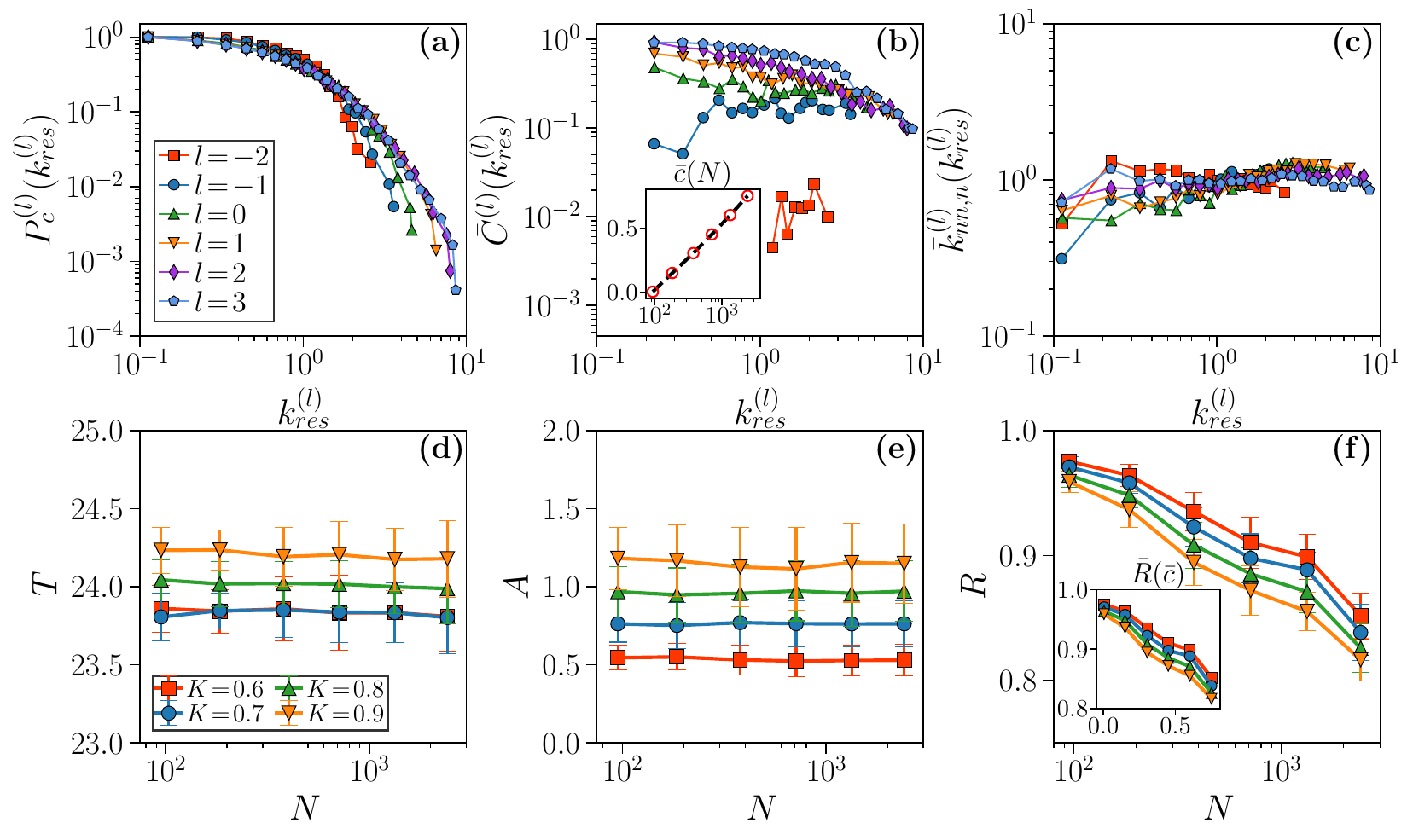}
\caption{
	\textbf{Impact of disrupting self-similarity in the clustering coefficient on circadian rhythms.}  
	(a) Complementary cumulative degree distribution.  
	(b) Degree-dependent clustering coefficient; inset: average clustering coefficient as a function of network size $N$, showing the expected increase with $N$.  
	(c) Degree--degree correlations.  
	In (a-c), degrees are rescaled as $k_{\mathrm{res}} = k / \bar{k}$.  
	(d) Average period $T$, (e) SCN amplitude $A$, and (f) synchronization degree $R$ as functions of $N$ in the perturbed networks from Null-$c$ model under varying coupling strengths. Error bars represent one standard deviation from the mean. Results are averaged over $50$ realizations with different initial conditions.
}
\label{fig:5}
\end{figure*}

\subsection{Impact of disrupting self-similarity in clustering on circadian rhythms}

While increasing average degree with network size induces size-dependent circadian rhythms, a natural question arises: how are biological rhythms affected when average clustering increases with network size? Such a scenario necessarily disrupts self-similarity, altering the balance of local and global connectivity and potentially reshaping the collective dynamics of circadian rhythms.

To explore the impact of disrupting network self-similarity on biological rhythms, we construct a \emph{Null-$c$} model, analogous to the Null-$k$ model, that increases average clustering with network size by edge swapping (see Appendix~\ref{SI-C} for more details). Specifically, starting from the scaled-up and scaled-down replicas, we adjust the targeted average clustering in each layer via edge swapping. The targeted clustering is artificial: in layer $l=-2$, it is reduced to $0$, and the clustering is then increased in artificial steps of $\Delta C = 0.15$ from $l=-2$ onward. As shown in Figs.~\ref{fig:5}(a-c), the Null-$c$ model preserves node degrees while disrupting self-similarity in clustering. 
As expected, the average clustering increases with network size $N$ (see Fig.~\ref{fig:5}(b), inset), and the curves of the degree-dependent clustering coefficient across layers no longer overlap, indicating a loss of scale invariance.

Circadian rhythms are insensitive to the loss of clustering self-similarity in the Null-$c$ model. As shown in Figs.~\ref{fig:5}(d) and~\ref{fig:5}(e), the average period and amplitude remain nearly unchanged with increasing network size $N$ for a given coupling strength. Disrupting clustering self-similarity causes the synchronization order parameter $R$ to decrease slightly with network size $N$ (Fig.~\ref{fig:5}(f)), or equivalently, $R$ decreases as the average clustering $\bar{c}$ increases (Fig.~\ref{fig:5}(f), inset). Nevertheless, strong synchronization is consistently maintained ($R>0.8$). This suggests that clustering self-similarity has only a minor role in regulating circadian rhythm robustness.

\section{Conclusions and discussion}

Studying multi-scale circadian rhythms in SCN networks is crucial for understanding how network structure shapes rhythmic robustness, sustains stable neural oscillator activity across scales, and reveals general principles linking network topology to biological timekeeping. Previous modeling studies of synthetic SCN networks revealed size-dependent circadian rhythms, with synchronization initially strengthening and then saturating~\cite{BWnt}. However, whether such size-dependent rhythms occur in real SCN networks remains unclear, as most studies focus on synthetic or single-scale empirical networks. Investigating these dynamics in multi-scale empirical SCN networks is therefore essential to uncover the structural determinants of robust circadian oscillations.

In this work, we applied the GBG and GR transformations to generate self-similar scaled-up and scaled-down replicas of five functional mouse SCN networks. We observed the absence of size-dependent circadian rhythms in these replicas, with the average period, amplitude, and synchronization degree remaining independent of network size. Using the Null-$k$ model, we reproduced size-dependent circadian rhythms and confirmed that they arise from the increase in average degree with network size. Networks with low average degree were fragmented into many components and isolated nodes, preventing oscillators from sustaining stable oscillations. Finally, disrupting clustering self-similarity slightly reduces synchronization, but circadian rhythms remain robust, with period and amplitude largely unaffected and strong synchronization maintained, suggesting that clustering self-similarity plays only a minor role in rhythm regulation.
 
Together, these findings highlight the robustness of SCN circadian rhythms to network scaling and provide new insights into how structural features shape biological rhythms. In particular, our results suggest that network average degree, rather than clustering self-similarity, is the dominant structural driver of synchronized circadian oscillations. This underscores the importance of degree-preserving mechanisms in the SCN, which may buffer biological rhythms against variations in scale or architecture.  

Several limitations of the present study should be acknowledged. First, our analysis relies on MIC-thresholded functional networks that capture statistical dependencies rather than direct anatomical connectivity; nevertheless, the GBG and GR methods remain robust across a range of reasonable threshold choices (Appendix~\ref{SI-Sensitivity}), supporting the stability of our conclusions. Second, the inferred networks are undirected, precluding causal or directional inference, making the incorporation of directionality an important direction for future work. Finally, the lack of cell-type--resolved information and spatial gradients of VIP and AVP expression prevents a direct biological interpretation of the observed hyperbolic maps or explicit modeling of localized neuronal clusters; integrating such data will be essential for more biologically grounded analyses.

Looking forward, our framework opens several promising directions for future research.
First, experimental validation using multiscale recordings of SCN activity could be used to test the predicted independence of circadian period and amplitude from network size.
Second, the perturbations introduced by the Null-$k$ and Null-$c$ models preserve a relatively homogeneous degree distribution and therefore do not aim to fully reproduce the known biological heterogeneity of the SCN.
In the current model, heterogeneous intrinsic periods are introduced by scaling system variables with factors sampled from a Gaussian distribution.
Investigating how more realistic forms of neuronal heterogeneity within biologically plausible parameter ranges influence circadian robustness thus represents an important direction for future work~\cite{wu2025neural}.
Finally, applying our methods to other brain networks with rhythmic functions may reveal general principles linking network topology, self-similarity, and biological rhythms.

\appendix

\section{Topological validation of hyperbolic embedding}
\label{SI-A}
\begin{figure*}[!ht]
	\centering
	\includegraphics[width=1\linewidth]{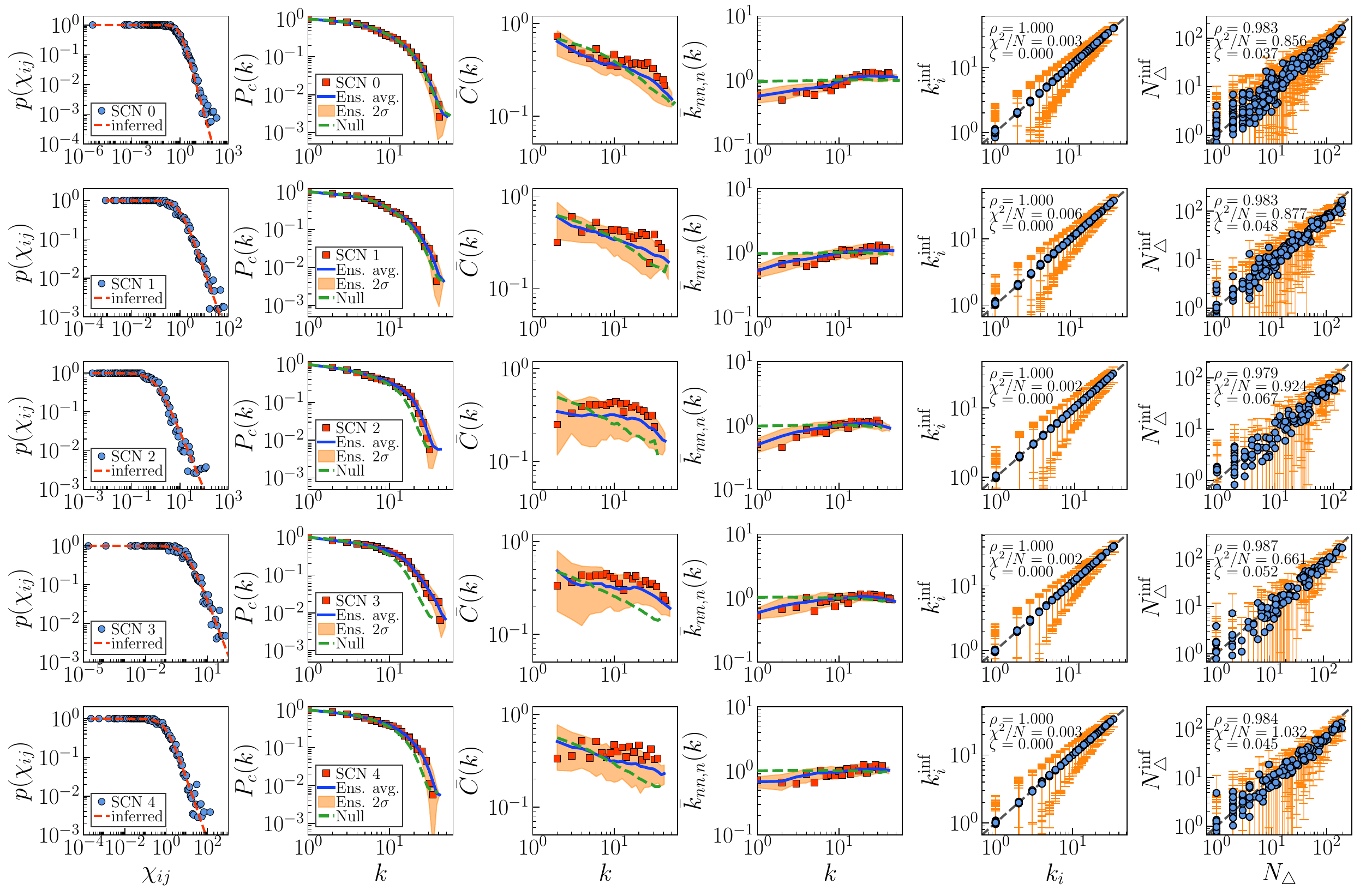}
\caption{\textbf{Topological validation of hyperbolic embedding.} 
	Each row corresponds to an empirical SCN network. 
	The first column compares the empirically measured connection probability with the theoretical prediction from Eq.~(\ref{eq:con_pro_S1}). 
	The second to fourth columns show the complementary cumulative degree distribution, degree-dependent clustering coefficient, and average nearest-neighbor degree, respectively. 
	Red symbols denote empirical measurements, blue solid lines show ensemble averages from the $\mathbb{S}^1$ model, orange shaded regions indicate the $2\sigma$ confidence intervals, and green dashed lines represent averages from degree-preserving null models.
	The fifth and sixth columns compare local node-level properties between the $\mathbb{S}^1$ model and empirical data, with degree shown in the fifth column and the number of triangles attached to each node in the sixth. 
	Error bars represent $2\sigma$ confidence intervals. 
	Statistical test results for $\rho$, $\chi^2$, and $\zeta$ are reported in each panel of the fifth and sixth columns.}
\label{fig:embedding}
\end{figure*}

\begin{figure*}[!ht]
	\centering
	\includegraphics[width=1\linewidth]{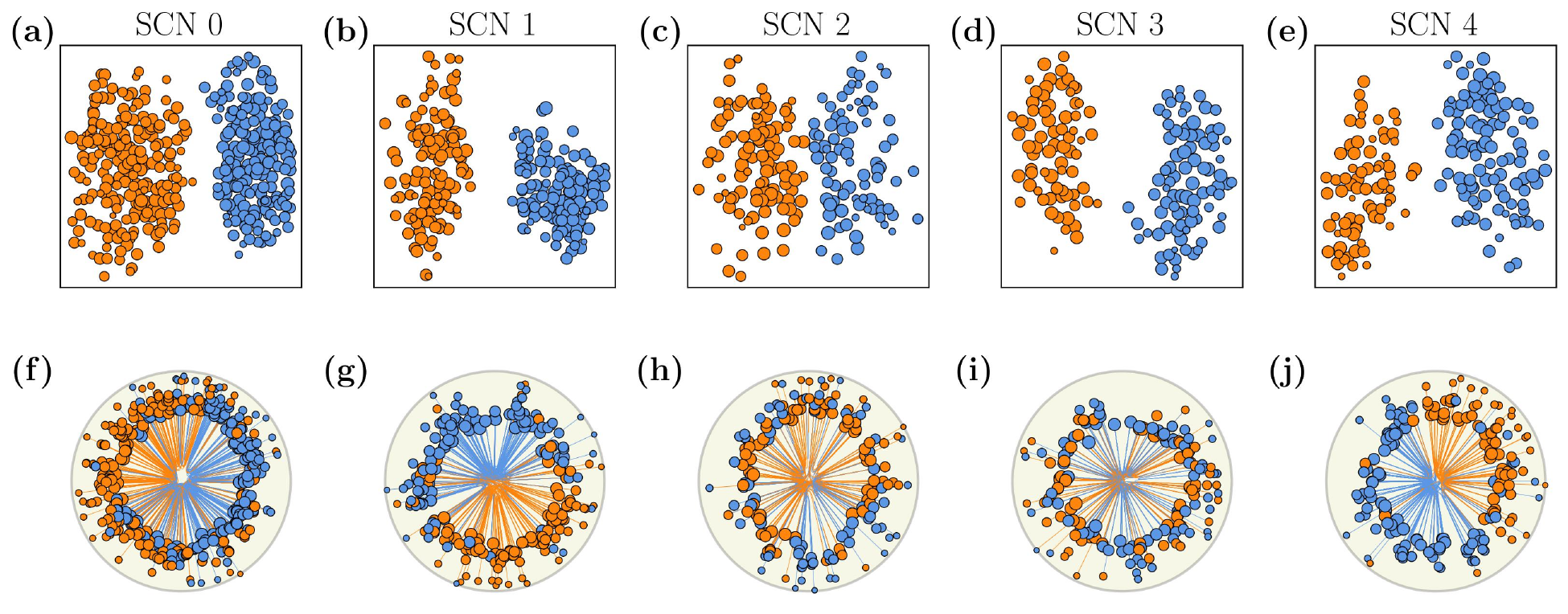}
	\caption{ {\bf Hyperbolic maps of SCN networks.} 
		Each column corresponds to an empirical SCN network.
		(a-e) Each SCN network is displayed in 2D Euclidean space. Node size is proportional to the logarithm of its degree, and node color indicates spatial location, with orange and blue representing the left and right SCN, respectively. For clarity, network links are not shown.
		(f-j) Hyperbolic embeddings of the corresponding SCN networks. Node sizes and colors are the same as in panels (a-e). Curves represent geodesics (shortest paths) between nodes, with line colors chosen randomly between the two nodes. }		
	\label{fig:map}
\end{figure*}

 \begin{figure*}[!ht]
 	\centering
 	\includegraphics[width=1\textwidth]{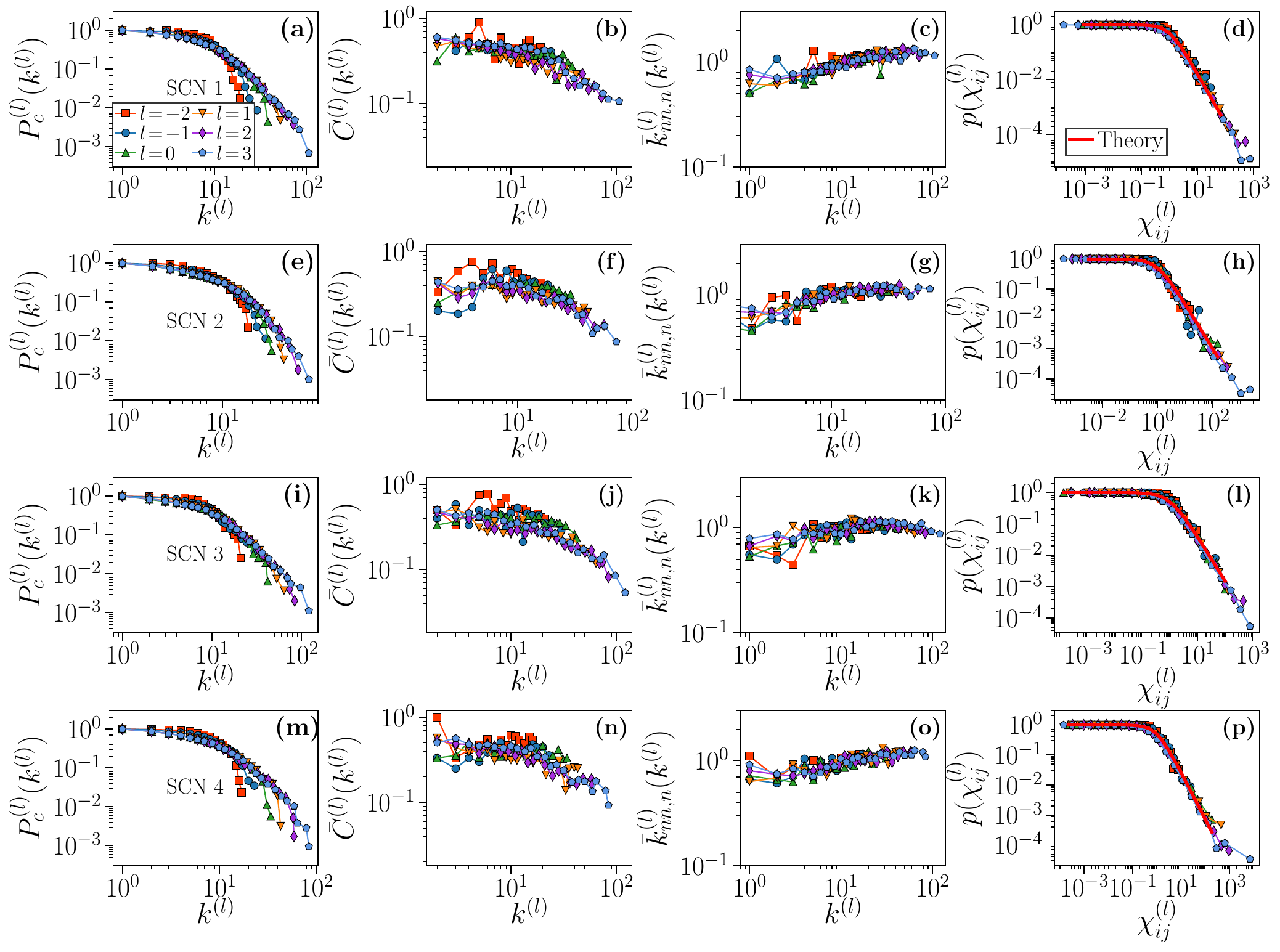}
 	\caption{\textbf{Topological properties of scaled-up and scaled-down replicas for the remaining SCN networks.}  
 		Each row corresponds to one empirical SCN network.  
 		The first column shows the complementary cumulative degree distribution, the second column the degree-dependent clustering coefficient, and the third column the degree--degree correlations.  
 		The fourth column presents the connection probability $p(\chi_{ij}^{(l)})$ as a function of the effective distance $\chi_{ij}^{(l)}$ in layer $l$. The red curve indicates the theoretical prediction from Eq.~\eqref{eq:con_pro_S1} for the original SCN network at $l=0$.}
 	\label{s2}
 \end{figure*}
We use the network embedding tool \textit{Mercator} to produce hyperbolic maps of SCN networks~\cite{Garcia2019}. The topological validation of the hyperbolic embedding for all SCN networks is shown in Fig.~\ref{fig:embedding}. 
In the first column, we compare the empirically measured connection probability with the theoretical prediction from Eq.~(\ref{eq:con_pro_S1}), showing excellent agreement between the empirical data and the model. 

To further assess the quality of the embeddings, we construct  $100$ synthetic networks as an ensemble using Eq.~(\ref{eq:con_pro_S1}) with the inferred $\{\kappa_i, \theta_i\}$ and parameters $\beta$ and $\mu$. The second to fourth columns in Fig.~\ref{fig:embedding} compare key topological properties between the synthetic ensemble and the empirical SCN networks: the complementary cumulative degree distribution $P_c(k)$, the degree-dependent clustering coefficient $\bar{C}(k)$, and the average nearest-neighbor degree $\bar{k}_{nn,n}(k)$, respectively. We also examined local node-level properties, shown in the fifth and sixth columns of Fig.~\ref{fig:embedding}, representing the degree and the number of triangles attached to each node. Additionally, we assessed the embeddings using statistical measures, including the Pearson correlation coefficient $\rho$, the normalized $\chi^2$ test, 
$
\chi^2 = \frac{1}{N}\sum_{i=1}^{N} \left(\frac{\text{value}_{\rm real} - \text{value}_{\rm ensemble}}{\sigma_{\rm ensemble}}\right)^2,
$ 
and the score $\zeta$, which quantifies the fraction of nodes whose values fall outside the $2\sigma$ confidence interval. 
In summary, the findings confirm that the $\mathbb{S}^1$ model replicates the key structural features of SCN networks.

Moreover, to provide a valuable benchmark for assessing the specificity of the embedding, we construct null models that preserve degree but not geometric structure. Specifically, we generate degree-preserving null models using Eq.~(\ref{eq:con_pro_S1}) with the inferred parameters $\beta$ and $\mu$, where the hidden degrees $\kappa_i$ are approximated by the observed degrees $k_i$. For each realization, the angular coordinates $\theta_i$ are randomly assigned in $[0,2\pi)$. This procedure preserves the degree sequence while removing geometric structure through the randomization of $\theta_i$. 
As shown in the second to fourth columns of Fig.~\ref{fig:embedding}, the degree-preserving null models fail to reproduce the observed topological structure when geometric constraints are not included (see the green dashed lines). This highlights the essential role of geometric organization in accurately capturing the structure of SCN networks.

Finally, the inferred hyperbolic maps provide a biologically meaningful representation of network organization. 
Figure~\ref{fig:map}(a-e) shows each SCN network in 2D Euclidean space, where node size reflects the logarithm of degree and node color indicates spatial location, with orange and blue denoting the left and right SCN. These panels reveal a spatial hierarchy in degree distribution, with low-degree nodes mainly in the shell region and high-degree nodes in the SCN core.
Figure~\ref{fig:map}(f-j) shows the corresponding hyperbolic embeddings, with node sizes and colors matching those in Fig.~\ref{fig:map}(a-e). In the hyperbolic disk, the left and right SCN separate naturally, and core and shell neurons are arranged from the center toward the periphery. Together, these results suggest that hyperbolic embeddings may help identify core and shell neurons in SCN networks.

\section{Supplementary figures for the remaining SCN networks}
\label{SI-D}

We present additional results in the supplementary figures to further support our findings. In Fig.~\ref{s2}, we show the topological properties of scaled-up and scaled-down replicas for the remaining SCN networks. The results are consistent with SCN~0, displaying clear statistical self-similarity: curves for the complementary cumulative degree distribution, degree-dependent clustering coefficient, degree--degree correlations, and connection probability from networks of different sizes naturally collapse onto a single profile. These observations demonstrate that all SCN networks are self-similar, with structural patterns and topological properties preserved across scales. They confirm that the GBG and GR transformations provide a robust and generalizable framework for multi-scale analysis of complex networks.

\begin{figure*}[!ht]
	\centering
	\includegraphics[width=1.0\linewidth]{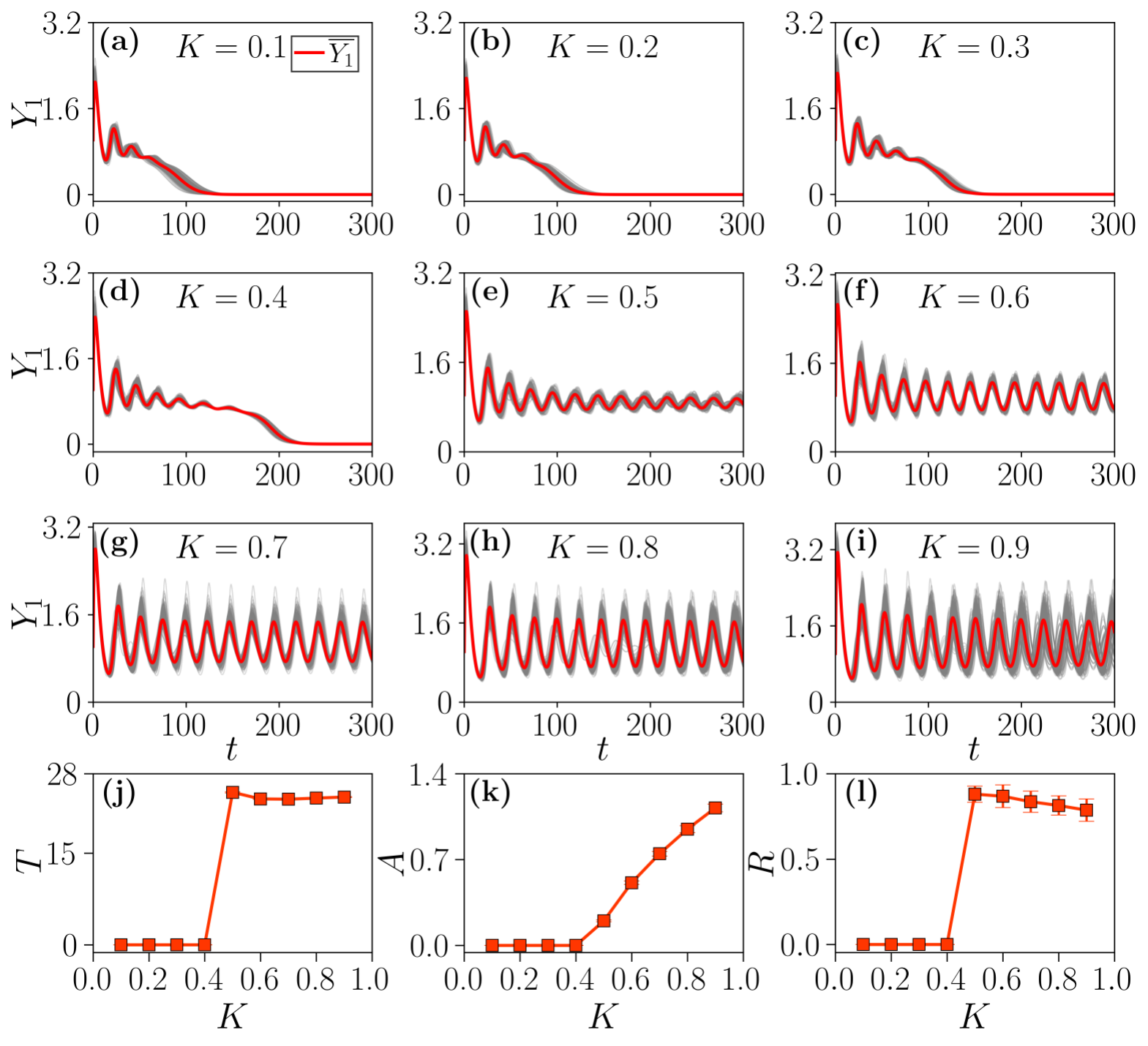}
	\caption{
		\textbf{Influence of coupling strength on circadian rhythms.}
		(a-i) Time series of the average output $\overline{Y_1}$ (red lines) and individual neuronal oscillators $Y_{1,i}$ (gray lines) in a representative SCN~0 network ($l=0$) for different coupling strengths $K$. 
		(j) Average period $T$, (k) SCN amplitude $A$, and (l) synchronization degree $R$ as functions of the coupling strength in the SCN~0 network. 
		Error bars represent one standard deviation from the mean. Results are averaged over $50$ realizations with different initial conditions and independently sampled $\mu_i$.
	}	
	\label{fig:K_check}
\end{figure*}

In Fig.~\ref{fig:K_check}, we illustrate the influence of the coupling strength $K$ on circadian rhythms in a representative SCN~0 network at $l=0$; qualitatively similar results are observed for the other SCN networks (not shown). Figures~\ref{fig:K_check}(a-i) display time series of the average output $\overline{Y_1}$ (red lines) together with individual neuronal oscillators $Y_{1,i}$ (gray lines) for increasing values of $K$. For weak coupling ($K \leq 0.4$), oscillations gradually decay and the system converges to a non-oscillatory steady state, corresponding to oscillation death. When $K$ exceeds a critical threshold ($K \approx 0.5$), sustained collective oscillations emerge, accompanied by increasing coherence among neurons. 
Panels (j-l) quantify this transition. The average period $T$ in panel (j) is effectively zero in the weak-coupling regime and undergoes a sharp transition to a circadian-scale value once oscillations are established. The SCN amplitude $A$ in panel (k) increases monotonically with $K$, reflecting enhanced coherent population-level output. The synchronization degree $R$ in panel (l) rises sharply near the onset of oscillations and then saturates at larger values of $K$.

These results highlight the distinct roles of coupling strength in regulating synchronization and amplitude. Once $K$ exceeds the phase-locking threshold, global synchronization rapidly saturates, rendering $R$ largely insensitive to further increases in $K$. In contrast, the oscillation amplitude $A$ remains sensitive to coupling strength, as stronger coupling suppresses phase dispersion and reduces destructive interference among oscillators, thereby enhancing coherent circadian rhythms. This separation of mechanisms explains why synchronization saturates while amplitude continues to increase in the strong-coupling regime.

\begin{figure*}[!ht]
	\centering
	\includegraphics[width=1\textwidth]{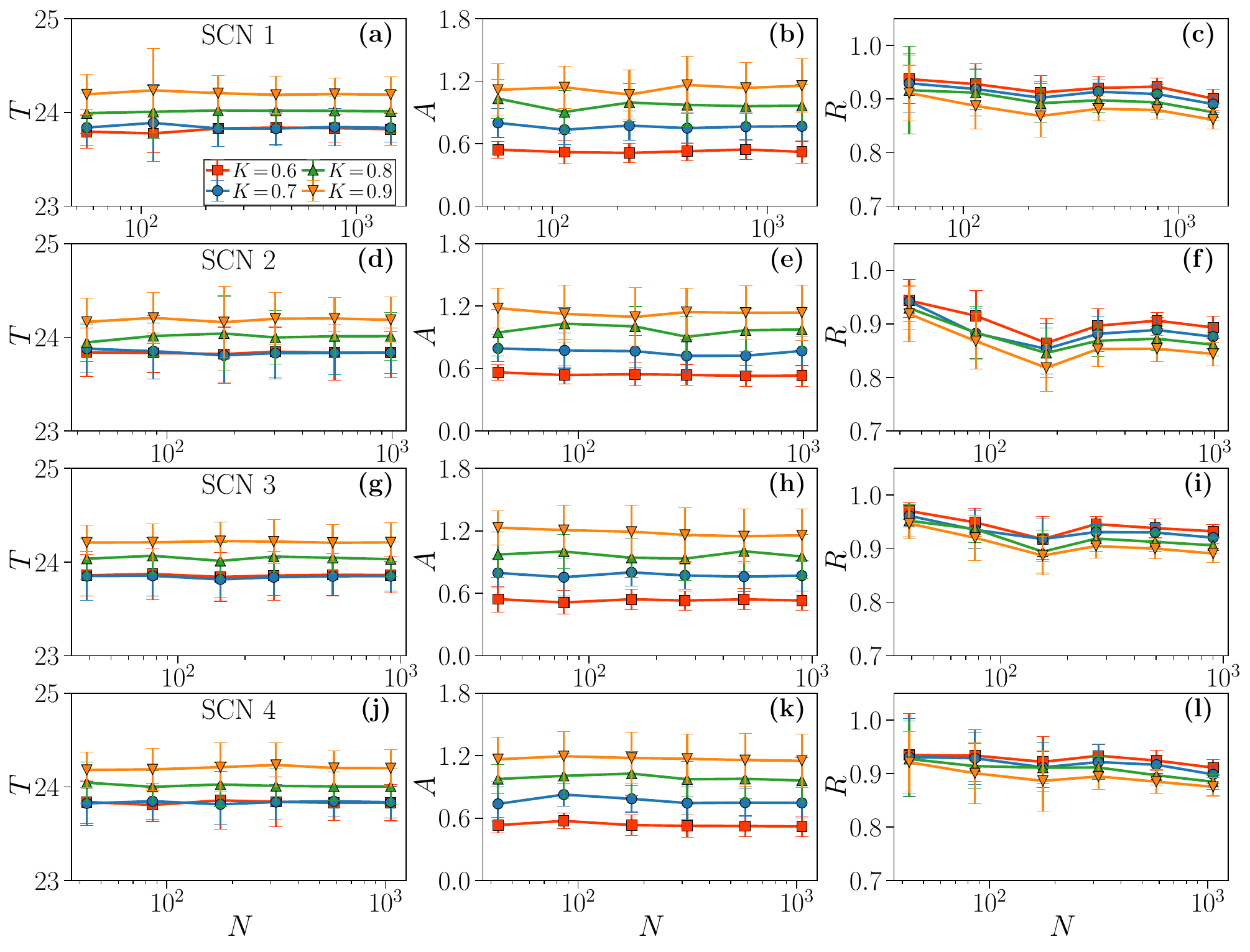}
	\caption{\textbf{Robustness of circadian rhythms in the remaining SCN replicas.}  
		Each row corresponds to one empirical SCN network.  
		From left to right, the columns show the average period $T$, amplitude $A$, and synchronization degree $R$ as functions of network size $N$ under varying coupling strengths.  
		Error bars denote one standard deviation from the mean.  
		All results are averaged over $50$ realizations with different initial conditions and independently sampled $\mu_i$.}
	\label{s3}
\end{figure*}

In Fig.~\ref{s3}, we examine the average period $T$, SCN amplitude $A$, and synchronization order parameter $R$ as functions of network size $N$ for the remaining SCN replicas. Circadian rhythms remain remarkably robust, with both the average period and amplitude nearly unchanged across scales. Strong synchronization is consistently observed ($R>0.80$) across all SCN networks, indicating coherent collective dynamics. These results confirm that circadian rhythms are stable and largely independent of network size in self-similar SCN replicas. Together, they provide further evidence that the multi-scale organization of SCN networks preserves essential temporal dynamics, highlighting the resilience of the system to structural scaling and reinforcing the validity of the GBG and GR transformations for analyzing multi-scale network behavior.

\section{Supplementary table for the KS test}
\label{KStest}
We employ the Kolmogorov--Smirnov (KS) test as a quantitative measure to compare degree distributions across network layers~\cite{simard2011computing}. This statistic captures deviations between empirical distributions, with smaller KS distance and corresponding large $p$-values indicating close agreement with the original network. As summarized in Table~\ref{tab:KS_test} in Appendix~\ref{KStest}, the degree distributions of scaled networks are largely indistinguishable from that of the original network, apart from finite-size effects observed in the smallest replica. Together, these results provide quantitative confirmation of the qualitative collapse observed in network statistics and reinforce the conclusion that the underlying network structure is robustly preserved across scales.
\begin{table*}[!ht]
	\centering	
	\setlength{\tabcolsep}{6pt}
	\caption{Kolmogorov--Smirnov (KS) comparison of degree distributions across layers. 
		For each SCN dataset (SCN~0--SCN~4), the table reports the KS distance and associated $p$-value for the comparison between layer $l$ and the original network ($l=0$).}
	\label{tab:KS_test}	
	\begin{tabular}{l cc cc cc cc cc}
		\toprule
		\multirow{2}{*}{Layer comparison}
		& \multicolumn{2}{c}{SCN~0} 
		& \multicolumn{2}{c}{SCN~1} 
		& \multicolumn{2}{c}{SCN~2} 
		& \multicolumn{2}{c}{SCN~3} 
		& \multicolumn{2}{c}{SCN~4} \\
		\cmidrule(lr){2-3}\cmidrule(lr){4-5}\cmidrule(lr){6-7}\cmidrule(lr){8-9}\cmidrule(lr){10-11}
		& KS dist. & $p$-value 
		& KS dist. & $p$-value 
		& KS dist. & $p$-value 
		& KS dist. & $p$-value 
		& KS dist. & $p$-value \\
		\midrule
		$l=-2$ vs $l=0$ & 0.15 & 0.06 & 0.23 & 0.01 & 0.11 & 0.64 & 0.30 & 0.00 & 0.24 & 0.02 \\
		$l=-1$ vs $l=0$ & 0.11 & 0.09 & 0.12 & 0.18 & 0.08 & 0.73 & 0.14 & 0.14 & 0.13 & 0.16 \\
		$l=1$ vs $l=0$  & 0.03 & 0.93 & 0.04 & 0.83 & 0.12 & 0.01 & 0.06 & 0.60 & 0.06 & 0.46 \\
		$l=2$ vs $l=0$  & 0.04 & 0.84 & 0.04 & 0.66 & 0.11 & 0.01 & 0.07 & 0.26 & 0.06 & 0.36 \\
		$l=3$ vs $l=0$  & 0.05 & 0.48 & 0.05 & 0.34 & 0.11 & 0.00 & 0.07 & 0.10 & 0.07 & 0.14 \\
		\bottomrule
	\end{tabular}	 
\end{table*}

\section{
Scale-invariant SCN rhythms under light--dark cycles and Kuramoto dynamics
}
\label{light-dark} 

We have demonstrated scale-invariant SCN rhythms using the Becker--Weimann model under constant-darkness conditions, showing that the average period, amplitude, and synchronization degree remain stable across network scales. A natural question is whether this scale-invariant rhythmic behavior persists under more biologically realistic environmental conditions, such as periodic light--dark cycles. To address this, we apply the Becker--Weimann model with light--dark forcing implemented as a clipped sine wave, with the light intensity set to $L_0 = 0.05$. The results for a representative SCN~0 network, shown in Fig.~\ref{fig:light_dark} in Appendix~\ref{light-dark}, demonstrate that circadian rhythms under light--dark cycles remain robust and independent of network size in self-similar SCN replicas. The results are robust for the rest SCN networks (not shown in this paper). These findings indicate that the observed scale invariance is not an artifact of constant-darkness conditions but instead reflects an intrinsic property of the underlying network structure.

\begin{figure}[!t]
	\centering
	\includegraphics[width=1.0\linewidth]{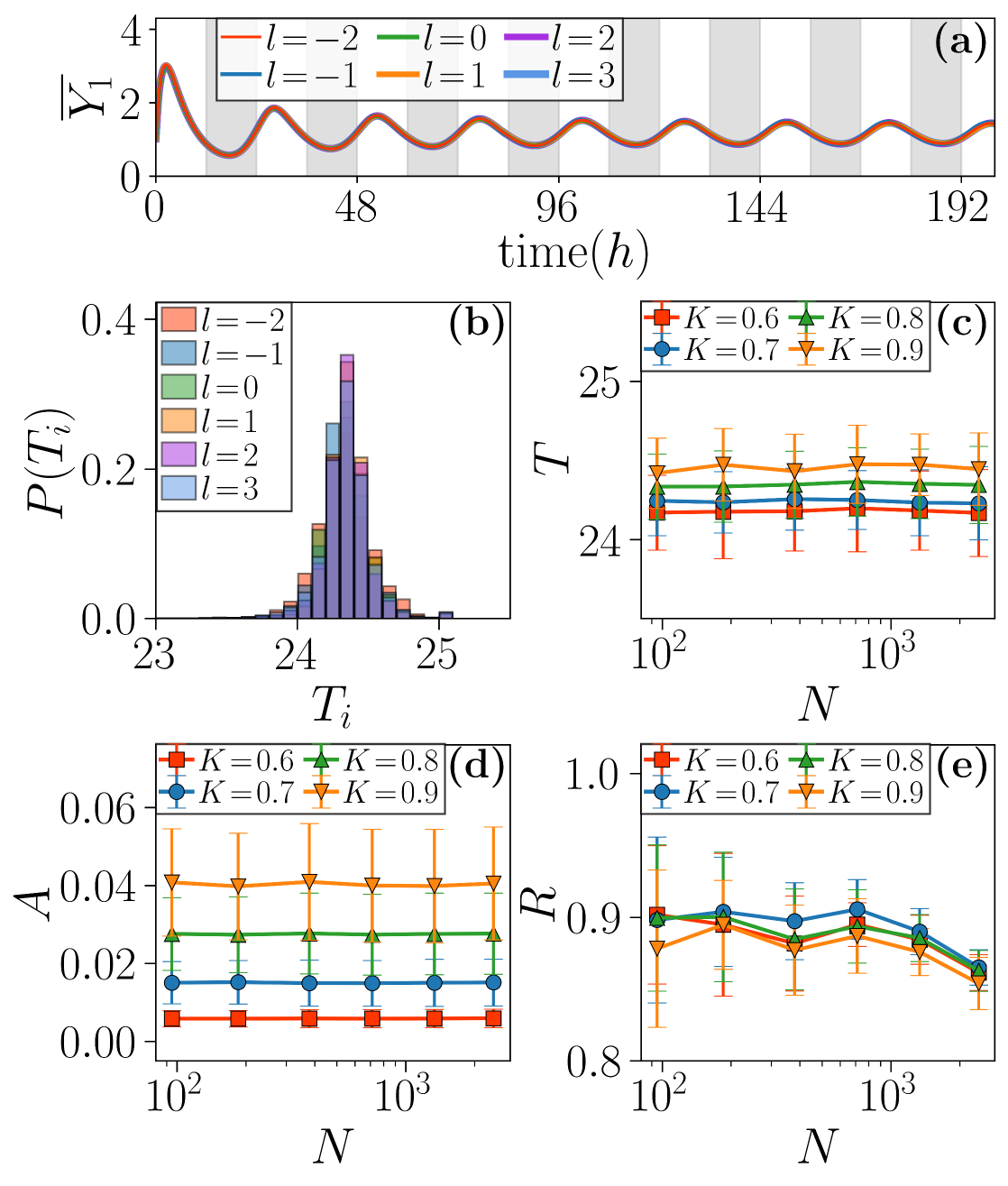}
	\caption{
		\textbf{Robustness of circadian rhythms in upscaled and downscaled SCN replicas under light--dark cycles.} Results are shown for a representative SCN~0 network and obtained using the Becker--Weimann model under light--dark forcing with light intensity $L_0 = 0.05$.
		(a) Time series of the average output $\overline{Y_1}$ across different layers $l$ for coupling strength $K = 0.8$.
		(b) Corresponding period distribution $P(T_i)$ of individual oscillators for $K = 0.8$.
		(c) Average period $T$, (d) SCN amplitude $A$, and (e) synchronization degree $R$ as functions of network size $N$ for different coupling strengths.
		Error bars represent one standard deviation from the mean. Results are averaged over $50$ realizations with different initial conditions and independently sampled $\mu_i$.
	}	
	\label{fig:light_dark}
\end{figure}

A natural next question is whether the scale-invariant rhythmic behavior observed above persists across different oscillator models. To address this, we examine SCN replicas using the Kuramoto model and analyze the average oscillation period and synchronization degree across network scales~\cite{moreno2004synchronization,gu2019heterogeneity}. 
The Kuramoto model is a widely used phase-based framework that captures the essential features of synchronization dynamics and has been extensively applied to the study of circadian systems.
The Kuramoto dynamics are governed by
\begin{equation}
	\dot{\phi}_i=\omega_i+K\sum_{j=1}^N
	a_{ij}\sin\!\left(\phi_j - \phi_i\right),
\end{equation}
where $\phi_i$ denotes the phase of oscillator $i$, $\omega_i$ is its intrinsic frequency, and $a_{ij}$ is the adjacency matrix element. The intrinsic frequencies $\omega_i$ are independently drawn from a normal distribution with mean $2\pi/24$ and standard deviation $0.05$, and the initial phases $\phi_i$ are uniformly distributed in the interval $[0,2\pi)$.

Here, $\phi_i$ is used as a marker of the dynamical evolution of individual neuronal oscillators. Accordingly, the circular mean phase of the oscillator population is given by the argument of the complex order parameter, 
\begin{equation}
	\overline{\phi} = \arg\!\left(\frac{1}{N}\sum_{i=1}^{N} e^{\mathrm{i}\phi_i}\right).
\end{equation}
The average oscillation period is defined as
\begin{equation}
	T = \frac{1}{N} \sum_{i=1}^{N} T_i,
\end{equation}
where $T_i$ denotes the oscillation period of the phase $\phi_i$.
To characterize collective synchronization, we introduce the synchronization degree
\begin{equation}
	R = \frac{1}{N} \left\langle \left| \sum_{j=1}^{N} e^{\mathrm{i} \phi_j} \right| \right\rangle ,
	\label{eq:synchronization}
\end{equation}
where $\langle \cdots \rangle$ represents a temporal average.
The system is fully desynchronized when $R = 0$, whereas $R = 1$ corresponds to perfect phase synchronization.

Numerical simulations are performed using the fourth-order Runge--Kutta method with a time step of $0.01 \mathrm{h}$. To eliminate transient effects and ensure convergence to stationary dynamics, the first $50000$ time steps ($500 \mathrm{h}$) are discarded prior to data analysis.

\begin{figure*}[!ht]
	\centering
	\includegraphics[width=1\linewidth]{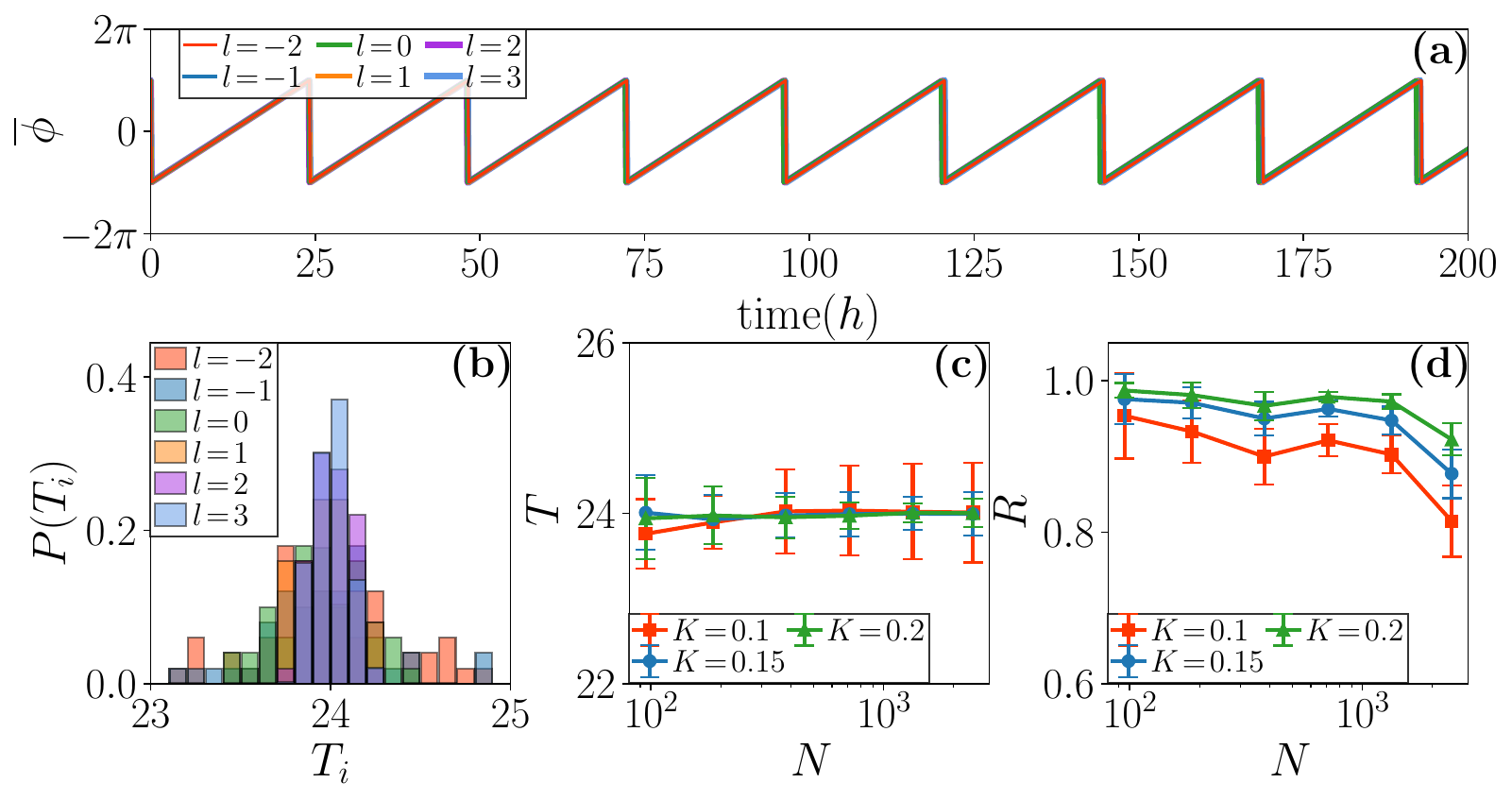}
	\caption{
		\textbf{Robustness of circadian rhythms in upscaled and downscaled SCN replicas within the Kuramoto model.}
		Results are shown for a representative SCN~0 network; similar behavior is observed for the other SCN networks (not shown).
		(a) Time series of the mean phase of the oscillator population, $\overline{\phi}$, across different layers $l$ for coupling strength $K=0.2$.
		(b) Corresponding distribution of individual oscillator periods, $P(T_i)$, for $K=0.2$.
		(c) Average oscillation period $T$ and (d) synchronization degree $R$ as functions of network size $N$ for different coupling strengths.
		Error bars represent one standard deviation across realizations.
		All results are averaged over $50$ realizations with different initial conditions.
		The observed robustness is consistent with that reported for the Becker--Weimann model.
	}
	\label{fig:Kuramoto}
\end{figure*}

The scale-invariant rhythmic behavior observed in the Becker--Weimann model persists when the dynamics are described by the Kuramoto model. 
As shown in Fig.~\ref{fig:Kuramoto}, circadian rhythms in self-similar SCN networks remain robust across scales within the Kuramoto framework. 
The mean phase $\overline{\phi}$ exhibits stable, periodic oscillations that closely overlap across different network layers, indicating preserved collective dynamics despite substantial changes in network size. 
The distributions of individual oscillator periods are narrowly centered around $\sim 24$~h and remain largely invariant across layers. 
Consistently, the average period $T$ shows little dependence on network size $N$ for different coupling strengths, while the synchronization degree $R$ remains high ($R \gtrsim 0.8$) across all scales. 
Together, these results demonstrate that both the collective period and synchronization degree are largely independent of network size in self-similar SCN replicas, indicating that scale-invariant rhythmicity arises from a generic, network-driven mechanism of collective synchronization rather than from model-specific biophysical details.

\section{Null-$k$ model}
\label{SI-B}
To investigate how the average degree influences circadian rhythms, we introduce a \emph{Null-$k$} model, which varies the average degree while preserving the underlying self-similar structure of the network. The idea is to couple changes in network size with controlled adjustments of the average degree by extending the geometric renormalization (GR) and geometric branch growth (GBG) methods.

In this construction, the empirical network at layer $l=0$ serves as the reference and remains unmodified. For the coarse-grained GR layers ($l=-1,-2,\dots$), the targeted average degree is reduced by selectively pruning edges during renormalization.  
Specifically, after obtaining a renormalized network via GR, the average degree must be reduced to the target value $\bar{k}_t$ we set artificially. Following~\cite{Garcia2018}, we adjust $\mu_{\mathrm{new}}^{(l)}$ so that 
$\bar{k}_{\mathrm{new}}^{(l)} = \bar{k_t}$ 
using 
$\mu_{\mathrm{new}}^{(l)} = \frac{\varepsilon \, \bar{k_t}}{\bar{k}^{(l)}} \, \mu^{(l)},$ 
with $\varepsilon = 1$ initially.  
For each $\varepsilon$, the network is pruned: if $\bar{k}_{\mathrm{new}}^{(l)} > \bar{k_t}$, we update $\varepsilon \to \varepsilon - 0.1 u$; if $\bar{k}_{\mathrm{new}}^{(l)} < \bar{k_t}$, $\varepsilon \to \varepsilon + 0.1 u$, where $u \sim U(0,1)$. The process stops once 
$|\bar{k}_{\mathrm{new}}^{(l)} - \bar{k_t}| < 0.05$.

In contrast, for the synthetic GBG layers ($l=1,2,\dots$), we increase the targeted average degree by inserting additional edges while preserving geometric constraints.  
Specifically, after obtaining an enlarged network via GBG, the average degree is adjusted to the target value $\bar{k_t}$ set artificially. To tune the average degree, we define 
$a = \varepsilon \, \bar{k_t} / \bar{k}^{(l)}$, 
where $\bar{k_t}$ is the target and $\bar{k}^{(l)}$ is the non-inflated average degree at layer $l$. Starting with $\varepsilon = 1$, we iteratively add links using Eq.~\eqref{eq:conprob_extra} and compute $\bar{k}_{\mathrm{new}}^{(l)}$. If $\bar{k}_{\mathrm{new}}^{(l)} > \bar{k_t}$, the realization is discarded and $\varepsilon \to \varepsilon - 0.1\,u$; if $\bar{k}_{\mathrm{new}}^{(l)} < \bar{k_t}$, then $\varepsilon \to \varepsilon + 0.1\,u$, where $u \sim U(0,1)$. The process terminates once 
$|\bar{k}_{\mathrm{new}}^{(l)} - \bar{k_t}| < 0.05$.

In this way, the Null-$k$ model allows us to disentangle the effects of connectivity density from those of system size. As an illustrative example, Fig.~\ref{fig:3} shows the structural properties of the resulting networks under a simple artificial scaling rule, 
$\overline{k}^{(l)} = 2 \, \overline{k}^{(l-1)}$,
where the average degree doubles at each layer.

\begin{figure*}[!ht]
	\centering
	\includegraphics[width=0.93\textwidth]{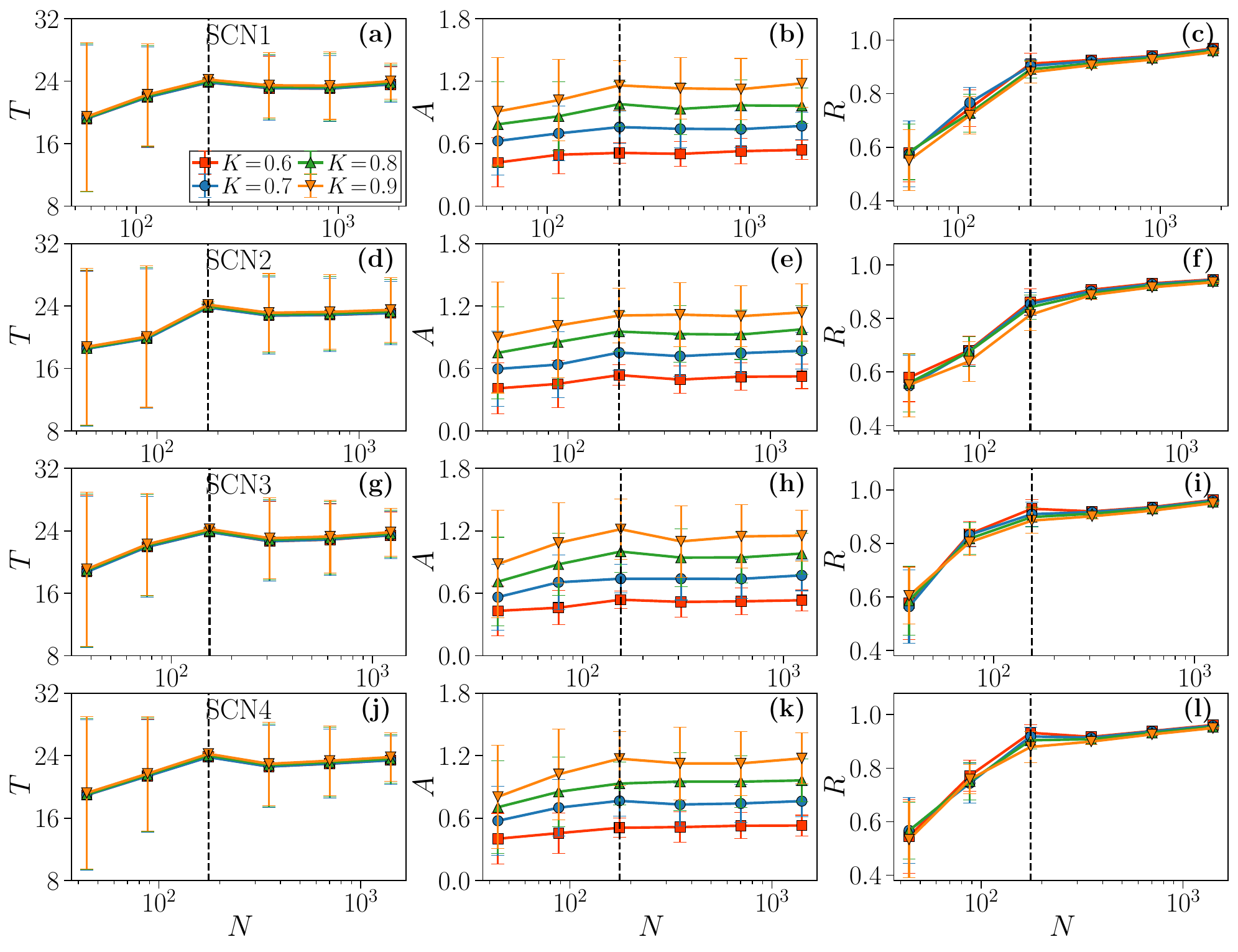}
	\caption{\textbf{Size-dependent circadian rhythms in the remaining SCN networks under the Null-$k$ model.}  
		Each row corresponds to one empirical SCN network.  
		From left to right, the columns show the average period $T$, amplitude $A$, and synchronization degree $R$ as functions of network size $N$ under varying coupling strengths.  
		Error bars indicate one standard deviation, and black dashed vertical lines mark the original networks.  
		Each data point is averaged over $50$ realizations with different initial conditions.}
	\label{s4}
\end{figure*}

In Fig.~\ref{s4}, we examine size-dependent circadian rhythms in networks generated by the Null-$k$ model for the remaining SCN networks. We observe that the average period, amplitude, and synchronization degree initially increase with network size and then saturate. Compared with the original networks at $l=0$ (black dashed vertical lines in Fig.~\ref{s4}), smaller networks with lower average degree display higher sensitivity and greater variance in their circadian dynamics. All SCN networks exhibit clear network-dependent rhythms, indicating that the observed size dependence is primarily driven by the average degree.

\section{Null-$c$ model}
\label{SI-C}

To investigate how disrupting network self-similarity affects biological rhythms, we construct a \emph{Null-$c$} model, in analogy to the Null-$k$ model, but designed to manipulate clustering rather than degree. In this model, the average clustering coefficient is systematically increased with network size through controlled edge-swapping operations. 

Starting from both scaled-up and scaled-down replicas, we tune the clustering in each layer to a target value $\bar{C}_t$ by performing edge swaps while preserving the degree sequence. The target clustering values are imposed artificially. As an illustrative example, in layer $l=-2$ we set $\bar{C}_t = 0$; from this baseline, the clustering is then progressively increased in discrete steps of $\Delta C = 0.15$ as $l$ increases (see Fig.~\ref{fig:5}).  

More specifically, consider a layer with average clustering $\bar{C}^{(l)}$ and a desired target $\bar{C}_t$ (with either $\bar{C}_t > \bar{C}^{(l)}$ or $\bar{C}_t < \bar{C}^{(l)}$). We randomly select two edges, $A\!-\!B$ and $C\!-\!D$, and attempt to rewire them as $A\!-\!D$ and $B\!-\!C$. The swap is accepted if it increases the clustering when $\bar{C}_t > \bar{C}^{(l)}$ (or decreases it when $\bar{C}_t < \bar{C}^{(l)}$). Self-loops and multiple edges are strictly forbidden. This process is repeated iteratively until the network’s average clustering reaches the prescribed target $\bar{C}_t$ within a tolerance of $0.01$.

In this way, the Null-$c$ model decouples clustering from other structural features, providing a controlled framework to examine its influence on collective rhythmic behavior. 

In fact, a degree-preserving randomization based on unrestricted edge swapping can also be used to probe structural self-similarity, in which pairs of edges are randomly rewired while prohibiting self-loops and multiple edges. This procedure substantially reduces the average clustering coefficient while leaving the degree distribution unchanged, leading to a pronounced suppression of self-similarity in the clustering coefficient.
However, although degree-preserving randomization effectively destroys clustering, it does not allow systematic control over its magnitude, whereas the Null-$c$ model enables controlled tuning of clustering across network layers. Therefore, we adopt the Null-$c$ model to systematically investigate the role of clustering in multiscale self-similarity and collective dynamics.


\section{Sensitivity analysis of MIC threshold selection on multiscale self-similarity}
\label{SI-Sensitivity}
\begin{figure}[!t]
	\centering
	\includegraphics[width=1.0\linewidth]{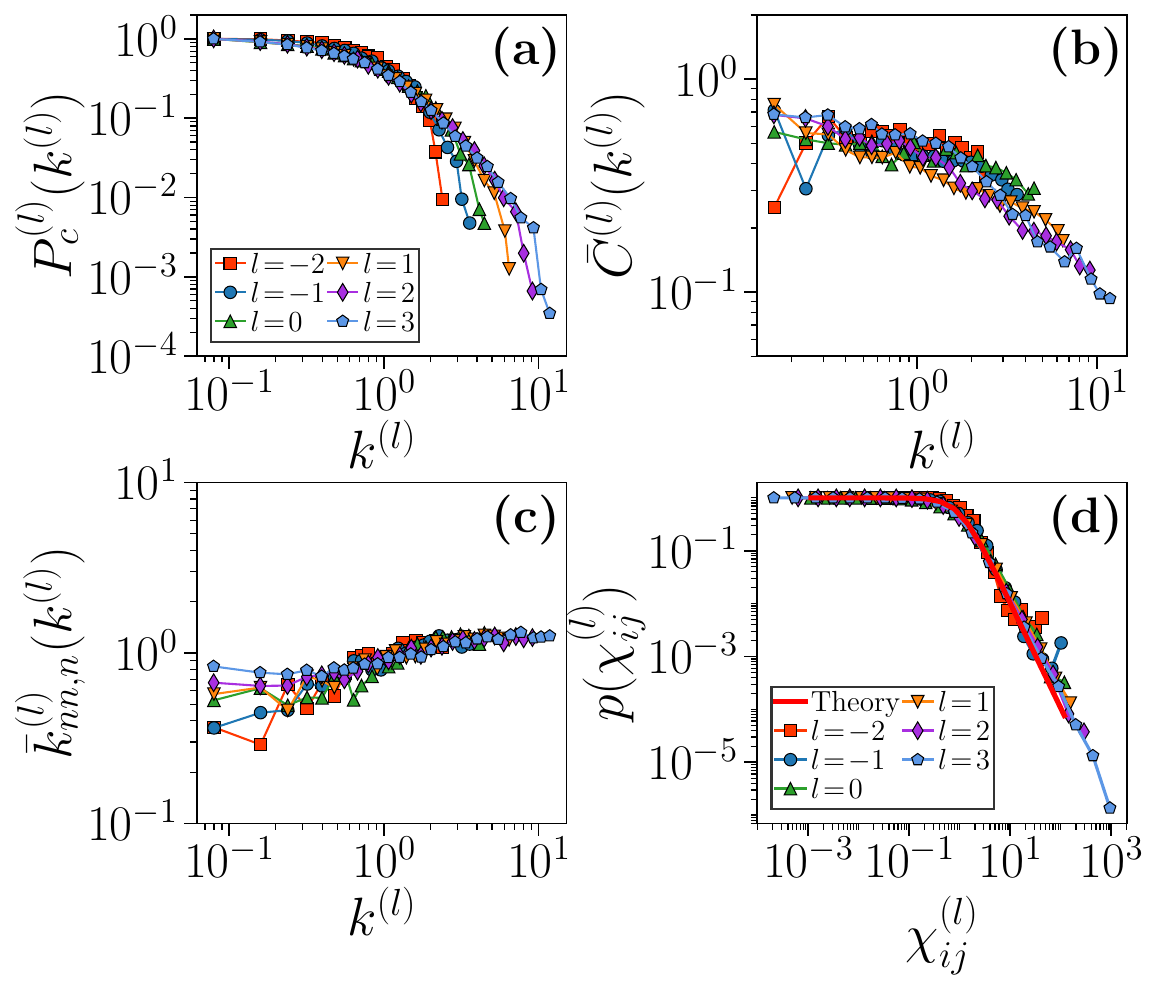}
	\caption{
		\textbf{Sensitivity analysis of self-similarity in scaled-up and scaled-down SCN~0 network constructed with an arbitrary MIC threshold.}
		The sensitivity analysis is performed on the starting snapshot of the SCN~0 network at $l=0$, constructed using an MIC threshold of $0.935$.
		(a) Complementary cumulative degree distribution.
		(b) Degree-dependent clustering coefficient.
		(c) Degree--degree correlations.
		(d) Connection probability $p(\chi_{ij}^{(l)})$ as a function of the effective distance $\chi_{ij}^{(l)}$ at layer $l$. The red curve indicates the theoretical prediction from Eq.~(\ref{eq:con_pro_S1}) for the starting snapshot at $l=0$.
	}
	\label{fig:MIC_check}
\end{figure}

Although the choice of MIC threshold can affect the inferred network topology by changing the number of retained connections and overall network density, the essential multiscale structural features remain stable over a broad range of reasonable thresholds. Consistent with this observation, the GBG and GR methods exhibit robust performance across networks constructed using different MIC thresholds. To illustrate this robustness, we apply both methods to a representative SCN~0 network constructed with an alternative MIC threshold of $0.935$. This network comprises $N = 422$ nodes, with an average degree of $12.53$ and an average clustering coefficient of $0.41$. The results are shown in Fig.~\ref{fig:MIC_check}.

\section*{Acknowledgments}
This work was supported by National Natural
Science Foundation of China (Grants No.~12305043, No.~12165016 and No.~12005079), the Natural Science Foundation of Jiangsu Province (Grant No.~BK20220511), 
M. Z. appreciates the support from the Jiangsu Specially-Appointed Professor Program.

%

\end{document}